\documentclass[article]{llncs}
\usepackage{fullpage}
\usepackage{amssymb}
\usepackage{graphicx}

\newcommand{\commentout}[1]{}

\begin{document}
\title{A Guide to Similarity Measures}

%%
%% The "author" command and its associated commands are used to define
%% the authors and their affiliations.
%% Of note is the shared affiliation of the first two authors, and the
%% "authornote" and "authornotemark" commands
%% used to denote shared contribution to the research.
\author{Avivit Levy \and B. Riva Shalom \and Michal Chalamish}
\institute{Shenkar College}

%%
%% By default, the full list of authors will be used in the page
%% headers. Often, this list is too long, and will overlap
%% other information printed in the page headers. This command allows
%% the author to define a more concise list
%% of authors' names for this purpose.
%\renewcommand{\shortauthors}{Levy et al.}

\commentout{
\author{
Avivit Levy\inst{1}
\and
B. Riva Shalom\inst{1}
\and
Michal Chalamish\inst{1}
}

\institute{Department of Software Engineering, Shenkar College, Ramat-Gan 52526, Israel.}

%\maketitle

%\setcounter{footnote}{0}
}

%\keywords{data science, similarity, distance, metric}

\maketitle
\begin{abstract}
Similarity measures play a central role in various data science application domains for a wide assortment of tasks. This guide describes a comprehensive set of prevalent similarity measures to serve both non-experts and professional. Non-experts that wish to understand the motivation for a measure as well as how to use it may find a friendly and detailed exposition of the formulas of the measures, whereas experts may find a glance to the principles of designing similarity measures and ideas for a better way to measure similarity for their desired task in a given application domain.
\end{abstract}
\section{Introduction}\label{s:introduction}
Similarity and distance measures play a central role in various data science applications for a wide assortment of tasks. Such applications include machine learning~\cite{roth2019mic,wu2017sampling}, artificial intelligence~\cite{rosenfeld2021better,ijcai2019-272,Amgoud_David_2021}, information retrieval~\cite{Jasiewicz2013,stewart2019information,ijcai2019-272}, text mining~\cite{Damerau:1964,winkler:1990,Amgoud_David_2021}, pattern matching~\cite{gu:2018,KPS:2021,clifford:2022}, pattern recognition~\cite{deng2019arcface,wang2019multi,WangWJHTZ21}, image processing~\cite{Maclem2002,WangWJHTZ21,qin2020deep}, computer vision~\cite{wang2021deep,tfw-2017,ijcai2019-417}, natural language processing~\cite{Lev:66,papineni2002bleu,Colombo2021}, networks security~\cite{Emran2002RobustnessOC,DeyMedya:2020}, data bases management~\cite{Jurman2009CanberraDO}, statistics~\cite{ChiSquared:1996,ChiSquared:2013}, computational biology~\cite{HillBurns2017,LiuHu:2019}, etc. Typical tasks that involve a use of similarity measures are correction~\cite{Lev:66}, reasoning~\cite{Amgoud_David_2021,Yu_Cao_Tang_Nie_Huang_Wu_2020}, prediction~\cite{stewart2019information}, correlation detection~\cite{correlation:2018}, recognition~\cite{taigman2014deepface,sun2014deep}, anomaly analysis~\cite{mahanabolis:2018}, clustering~\cite{noorbehbahani2015incremental} and classification~\cite{lwCanberra:66,Lorentzian:2016}.

The choice of a similarity or distance measure adequate for a specific task within an application domain is of great importance. For example, compressed image retrieval uses similarity and distance measures for evaluations, where some commonly used distance measures, as the Euclidean distance, do not give good retrieval performance, while others, such as the Canberra and Wave-Hedges metrics seem to be beneficial to retrieval efficiency and the computational complexity~\cite{HatzSkod2003}. 

Despite the critical impact that the choice of an adequate measure has, many researchers are unaware of the variety of possibilities to measure the similarity for a specific task in their domain of interest. Moreover, some better options may evade their knowledge scope. For example, the Mahalanobis distance~\cite{mahalanobis:1999} is an extremely useful metric having excellent applications in multivariate anomaly detection, classification on highly imbalanced data-sets and one-class classification. However, it is not so well known or used in the machine learning practice~\cite{mahanabolis:2018}.

We, therefore, intend to describe a comprehensive set of prevalent similarity measures used in various data science application domains for a wide assortment of tasks. Our description is meant to guide both non-experts and professionals. Non-experts that wish to understand the motivation for the measure as well as how to use it may find a friendly and detailed exposition of the measures formulas. Experts may find a glance to the principles of designing similarity measures and ideas for a better way to measure similarity in their desired task in a given application domain.  

In order to make our description comprehensive and to present a refined classification of similarity and distance measures, we considered several surveys and books (such as~\cite{Cha-Srihari:2002,c:2007,AL:10,Duda:2001,Deza:2006,deza2009encyclopedia}) as well as scanned overall about 80 papers presenting the relevant measures.  

Unlike the above mentioned surveys and books, our purpose in the presentation of the measures was different. On the one hand, we intended to include in this guide measures used for \emph{various data types}, both numerical and categorical, and for \emph{diverse data science applications}. We did not narrow our focus to measures designed for a specific data type in order to enable a potential reader to broaden the view on the subject, thus potentially find solutions that were out of a specific domain knowledge scope. On the other hand, we wanted a concise measures selection, presenting the principles as well as the diversity of possibilities for similarity measures design.

Table~\ref{tab:summery} details the similarity or distance measures scanned in this survey, categorized by  measures families.
\paragraph{\textbf{Similarity Measures and Metrics.}} Similarity and distance measures are closely related. In a similarity measure, the score is higher as the compared objects are more similar, whereas in a related distance measure the score is diminished as the compared objects are more similar, and typically has zero score if they are identical. In the mathematics literature, a non-negative distance function $d$ on a set $\mathrm{X}$, $d:\mathrm{X}\times\mathrm{X}\rightarrow \mathbb{R}$, is called a \emph{metric} if for all $x,y,z\in \mathrm{X}$ it satisfies the following three conditions: (1) Identity of Indiscernibles: $d(x,y)=0$ if and only if $x=y$, (2) Symmetry: $d(x,y)=d(y,x)$, (3) Triangle Inequality: $d(x,y)\leq d(x,z)+d(z,y)$. Though these conditions make a metric more powerful and desirable for use, many widespread distance measures in various data science applications fail to fulfill some of these conditions. Since this guide is oriented by data science applications needs and practice, we refer to common distance measures regardless of their being a metric or not. We, nevertheless, remark on this fact while trying to identify the metric conditions which they fail to fulfill, in such a case.  

%The main goal of this survey is to give a comprehensive guide to the variety of similarity and distance measures applicable in diverse tasks on various data types in data science applications.
%\paragraph{\textbf{Paper Goals.}} (2) To describe the principles of similarity and distance measures design for various data types and tasks.  

%\paragraph{\textbf{Paper Organization.}} Section~\ref{s:SimClass} describes a comprehensive set of prevalent similarity and distance measures as well as their detailed and meaningful classification.  Section~\ref{s:conclusion} concludes the paper.
\begin{table}[h]
    \centering
 \small{
 \begin{tabular}{|c|c|c|c|c|c|c|c|}
    \hline
       Inner  & Minkowski & Intersection & Entropy& $\chi ^2$ Family& Fidelity  & String   & String \\ 
      Product&   &   &  &  &  (Squared-Chord) &  Rearrangement  & Similarity\\ 
    \hline
     \hline
   \footnotesize{Subsec.~\ref{ss:InnerProd} }& \footnotesize{Subsec. \ref{ss:Minkowski}}& \footnotesize{Subsec.~\ref{ss:intersection}} &\footnotesize{Subsec.~\ref{ss:entropy}} & \footnotesize{Subsec.~\ref{ss:ChiSquared}} & \footnotesize{Subsec.~\ref{ss:Fidelity}}& \footnotesize{Subsec.~\ref{ss:Strings}}& \footnotesize{Subsec.~\ref{ss:Strings}}  \\
    \hline
    \hline
    \footnotesize{Inner Product} & Euclidean & Intersection & \footnotesize{Kullback-Leibler}& Pearson & Fidelity  & Hamming   & LCS \\ 
     Cosine & (Euclidean)$^2$  & Wave Hedges & J-Divergence &  Neyman & Bhattacharyya  & Levenshtein   & Jaro \\ 
      Angular & $L_1$ & S{\o}rensen & K-Divergence& Add. Sym.$\chi ^2$ & Hellinger   & Swap   & String \\ 
      Jaccard & $L_p$ & Kulczynski & Tops{\o}e& Spearman & Matusia  & Interchange   & N-Grams  \\ 
      Dice & $L_\infty$ & Jaccard & \footnotesize{Jensen-Shannon} & Squared $\chi ^2$ & Squared-Chord   & \footnotesize{Parallel-Interchange}   &  \\ 
       &   &   & Jensen Diff. & Divergence&  &    &  \\ 
      &  &  & SED& Clark&    &    &  \\ 
      &  &  & & Mahalanobis &    &    &  \\ 
      \hline
\end{tabular} }
   \caption{Similarity or distance measures appearing in this survey, categorized by measures families.}
    \label{tab:summery}
\end{table}

% end commentout

\section{Similarity Measures Description and Classification}\label{s:SimClass}
This section is devoted to describe and classify a comprehensive set of similarity measures. Our classification is based on that of~\cite{c:2007}, however, we significantly broadened it in order to refer to various data types, as well as refined it by including or, in some cases, omitting some measures. In addition, we presented the measures differently, including not only the formula for computing each measure accompanied by an example, but also an explanation of its design purpose. We grouped the measures by families and presented also some basic known variants of some measures. Overall, more than 50 similarity/distance measures and their notable variants are presented in section. We next describe the classification, where each measures class is presented in a separate subsection.
\subsection{Inner Product Based Measures}\label{ss:InnerProd} 
A basic way to measure similarity between data instances is to refer to them as vectors in a vector space over $\mathbb{R}$ and make use of an \emph{inner product}, defined as follows~\cite{deza2009encyclopedia}:
\begin{definition}\label{d:ip} {\rm {\bf [Inner Product]}}
Let $V$ be a vector space over $\mathbb{R}$.\footnote{In general, inner product can be defined also over the field $\mathbb{C}$. In this paper, we focus on the more frequent vector spaces in data science.} An \emph{inner product} $\langle,\rangle$ is a binary operation $V\times V \rightarrow \mathbb{R}$ with the following properties: (1)\emph{Positive definiteness:} $\forall P\in V$, $\langle P,P\rangle\geq 0$, and $\langle P,P\rangle=0$ if and only if $P=0$. (2) \emph{Symmetry:} $\forall P,Q\in V$, it holds that $\langle P,Q\rangle=\langle Q,P\rangle$. (3) {Bi-linearity:} $\forall P_1,P_2,Q\in V$ and $r_1,r_2\in\mathbb{R}$, it holds that: 1. $\langle P_1+P_2,Q\rangle=\langle P_1,Q\rangle+\langle P_2,Q\rangle$. 2. $\langle r_1 P_1+r_2 P_2,Q\rangle=r_1\langle P_1,Q\rangle+r_2\langle P_2,Q\rangle$.
\end{definition}
Inner product spaces generalize Euclidean vector spaces, the real $d$-space $\mathbb{R}^d$, in which the inner product is the \emph{dot} or \emph{scalar product} of Cartesian coordinates: $\langle P,Q\rangle=P\cdot Q=P^TQ=\sum_{i=0}^{d}P_i\cdot Q_i$. Some other inner product spaces examples are: (a) The vector space $\mathcal{M}_n$ of \emph{$n\times n$ real matrices} with the trace function, i.e., if $P$ and $Q$ are vectors in $M_n$, that is, they are $n\times n$ real matrices, then $\langle P, Q\rangle=tr(P^TQ)=\sum_{i=1}^{n}P^TQ[i,i]$. (b) The vector space $P_n$ of \emph{all polynomials of degree at most $n$}, with standard inner product on $P_n$ defined by $\langle P,Q\rangle=\sum_{i=0}^{n}P_i\cdot Q_i$. (c) The vector space of \emph{real random variables} with the expected value of their product as the inner product, i.e,\ $\langle P,Q\rangle=\mathbf{Exp}[P\cdot Q]$.

\paragraph{\textbf{Norm in Inner Product Spaces:}} An inner product space induces a \emph{norm}: $\|P\|=\sqrt {\langle P,P\rangle }$.

Below, we describe similarity measures on vector spaces based on the inner product operation.
\begin{enumerate}
\item \textbf{Inner Product Distance and Similarity.}
With the above norm, every inner product space becomes a \emph{metric space}, with the distance defined by~\cite{deza2009encyclopedia}:
$$d_{IP}(P,Q)=\|P-Q\|$$

The Inner Product Similarity is the following:
$$sim_{IP}(P,Q)=\langle P,Q\rangle$$
For binary vectors over $\{0,1\}^d$ it measures the \emph{number of matches} or \emph{overlap} of vectors~\cite{c:2007}. We will use the following example throughout this subsection. 
Let $P=\left(
                      \begin{array}{cc}
                        1 & 2 \\
                        3 & 4 \\
                      \end{array}
                    \right)$ and $Q=\left(
                                      \begin{array}{cc}
                                        -1 & -2 \\
                                        -3 & -4 \\
                                      \end{array}
                                    \right)$ be vectors in the vector space $\mathcal{M}_2$ of \emph{$2\times 2$ real matrices} with the trace function as their inner product. Then,
$$sim_{IP}(P,Q)=tr(P^T Q)=\sum_{i=1}^{2}(P^T Q)[i,i]=\sum_{i=1}^{2}\left(
                           \begin{array}{cc}
                                        -10 & -14 \\
                                        -14 & -20 \\
                           \end{array}
                        \right) [i,i]=-30$$
and 
\commentout{
\begin{eqnarray}
% \nonumber to remove numbering (before each equation)
\nonumber  d_{IP}(P,Q) &=& \sqrt{tr((P-Q)^T (P-Q))}=\sqrt{tr(\left(
                           \begin{array}{cc}
                                        2 & 4 \\
                                        6 & 8 \\
                           \end{array}
                        \right)^T \left(
                           \begin{array}{cc}
                                        2 & 4 \\
                                        6 & 8 \\
                           \end{array}
                        \right))}= \\
\nonumber   &=&\sqrt{\sum_{i=1}^{2}\left(
                           \begin{array}{cc}
                                        40 & 56 \\
                                        40 & 80 \\
                           \end{array}
                        \right) [i,i]}=\sqrt{120}=10.95 
\end{eqnarray}
}
\begin{eqnarray}
% \nonumber to remove numbering (before each equation)
\nonumber  d_{IP}(P,Q) &=& \sqrt{tr((P-Q)^T (P-Q))}=\sqrt{tr(\left(
                           \begin{array}{cc}
                                        2 & 4 \\
                                        6 & 8 \\
                           \end{array}
                        \right)^T \left(
                           \begin{array}{cc}
                                        2 & 4 \\
                                        6 & 8 \\
                           \end{array}
                        \right))}= \sqrt{120}=10.95 
\end{eqnarray}

\item \textbf{Cosine Similarity.}
When $\langle P,Q\rangle$ is a real number then the Cauchy–Schwarz inequality, i.e.,\ $|\langle P,Q\rangle |\leq \|P\|\,\|Q\|$, guarantees that $\frac{\langle P,Q\rangle}{\|P\|\,\|Q\|}\in [-1,1]$ as the cosine trigonometric function, so the cosine similarity between $P$ and $Q$ is defined as~\cite{deza2009encyclopedia}:
$$sim_{\mathit{Cos}}(P,Q)=\frac{\langle P,Q\rangle}{\|P\|\,\|Q\|}$$

The geometric interpretation of this measure is the cosine of the angle between two vectors defined using an inner product. The term cosine distance is commonly used for the complement of cosine similarity in positive space, that is:
$$d_{\mathit{Cos}}(P,Q)=1-sim_{\mathit{Cos}}(P,Q)$$

For example, let $P$ and $Q$ be the above example vectors in the vector space $\mathcal{M}_2$ of \emph{$2\times 2$ real matrices} with the trace function as their inner product. We have that:
$$\|P\|=\sqrt{tr(\left(\begin{array}{cc}
                                        1 & 2 \\
                                        3 & 4 \\
                           \end{array}\right)^T \left(
                           \begin{array}{cc}
                                        1 & 2 \\
                                        3 & 4 \\
                           \end{array}\right))}= \sqrt{tr(\left(
                           \begin{array}{cc}
                                        10 & 14 \\
                                        14 & 20 \\
                           \end{array}\right)}=\sqrt{30}$$
$$\|Q\|=\sqrt{tr(\left(\begin{array}{cc}
                                        -1 & -2 \\
                                        -3 & -4 \\
                           \end{array}\right)^T \left(
                           \begin{array}{cc}
                                        -1 & -2 \\
                                        -3 & -4 \\
                           \end{array}\right))}= \sqrt{tr(\left(
                           \begin{array}{cc}
                                        10 & 14 \\
                                        14 & 20 \\
                           \end{array}\right)}=\sqrt{30}$$

Then,
$$sim_{\mathit{Cos}}(P,Q)=\frac{\langle P,Q\rangle}{\|P\|\,\|Q\|}=\frac{-30}{\sqrt{30}\sqrt{30}}=-1$$
and 
$$d_{\mathit{Cos}}(P,Q)=1-sim_{\mathit{Cos}}(P,Q)=2$$
Note, however, that the cosine distance is not a metric as it violates the triangle inequality.

\item \textbf{Angular Distance.}
Since $\frac{\langle P,Q\rangle}{\|P\|\,\|Q\|}\in [-1,1]$ is within the domain of the inverse trigonometric function $\arccos :[-1,1]\to [0,\pi ]$, the (non-oriented) angle between $P$ and $Q$ can be defined as: $\arccos{\frac{\langle P,Q\rangle}{\|P\|\,\|Q\|}}$, where $0\leq \arccos{\frac{\langle P,Q\rangle}{\|P\|\,\|Q\|}}\leq \pi$.

The normalized angle, referred to as angular distance, between any two vectors $P$ and $Q$ is a formal metric and can be calculated from the cosine similarity, as follows~\cite{deza2009encyclopedia}:
$$d_{\mathit{Ang}}(P,Q)=\frac{\arccos{sim_{\mathit{Cos}}}}{\pi}=\frac{\arccos{\frac{\langle P,Q\rangle}{\|P\|\,\|Q\|}}}{\pi}$$

The complement of the angular distance metric can then be used to define angular similarity function bounded between 0 and 1, inclusive:
$$sim_{\mathit{Ang}}(P,Q)=1-d_{\mathit{Ang}}(P,Q)$$
The geometric interpretation is the angle between two vectors.

For example, let $P$ and $Q$ be the above example vectors in the vector space $\mathcal{M}_2$ of \emph{$2\times 2$ real matrices} with the trace function as their inner product. Then,
$$d_{\mathit{Ang}}(P,Q)=\frac{\arccos{sim_{\mathit{Cos}}}}{\pi}=\frac{\arccos(-1)}{\pi}=\frac{\pi}{\pi}=1$$
and
$$sim_{\mathit{Ang}}(P,Q)=1-d_{\mathit{Ang}}(P,Q)=0$$

Unfortunately, computing the arc-cosine function is rather slow, making the use of the angular distance more computationally expensive than using the cosine distance.

\item \textbf{Jaccard Coefficient or Jaccard Index.}\footnote{Also called, Tanimoto coefficient/index~\cite{tani:58}.}
The Jaccard similarity is a variation of a normalized inner product usually applied to measure the similarity and diversity of sample sets by taking the size of overlap between the sets divided by their union size. By inner product means, it is formally defined as follows~\cite{Jaccard}:
$$sim_{\mathit{Jac}}(P,Q)=\frac{\langle P,Q\rangle}{{\|P\|}^2+{\|Q\|}^2-\langle P,Q\rangle}$$

The Jaccard distance applied to sets measures the size of the difference between the sets divided by their union size, and by inner product means is then formally defined:
$$d_{\mathit{Jac}}(P,Q)=1-sim_{\mathit{Jac}}(P,Q)=\frac{\|P-Q\|^2}{{\|P\|}^2+{\|Q\|}^2-\langle P,Q\rangle}$$
The Jaccard index is a formal metric.

For example, let $P$ and $Q$ be the above example vectors in the vector space $\mathcal{M}_2$ of \emph{$2\times 2$ real matrices} with the trace function as their inner product. Then,
$$sim_{\mathit{Jac}}(P,Q)=\frac{-30}{{\sqrt{30}}^2+{\sqrt{30}}^2-(-30)}=\frac{-30}{90}=-\frac{1}{3}$$
and
$$d_{\mathit{Jac}}(P,Q)=1-sim_{\mathit{Jac}}(P,Q)=1\frac{1}{3}$$

\item \textbf{Dice Coefficient.}\footnote{Also called, S{\o}rensen distance~\cite{sorensen:1948}.}
The Dice similarity~\cite{Dice:1945} is another variation of a normalized inner product applied to measure the similarity and diversity of sample sets by taking the size of overlap between the sets divided by their average size. By inner product means it is defined:
$$sim_{\mathit{Dice}}(P,Q)=\frac{2\langle P,Q\rangle}{{\|P\|}^2+{\|Q\|}^2}$$

The Dice distance is then defined:
$$d_{\mathit{Dice}}(P,Q)=1-sim_{\mathit{Dice}}(P,Q)=\frac{\|P-Q\|^2}{{\|P\|}^2+{\|Q\|}^2}$$
Unlike Jaccard, the Dice coefficient is not a metric as it violates the triangle inequality.

For example, let $P$ and $Q$ be the above example vectors in the vector space $\mathcal{M}_2$ of \emph{$2\times 2$ real matrices} with the trace function as their inner product. Then,
$$sim_{\mathit{Dice}}(P,Q)=\frac{2(-30)}{{\sqrt{30}}^2+{\sqrt{30}}^2}=-1$$
and
$$d_{\mathit{Dice}}(P,Q)=1-sim_{\mathit{Dice}}(P,Q)=2$$
\end{enumerate}

\subsection{The Minkowski Distance Family}\label{ss:Minkowski}
The measures described in this subsection share a basic formula structure, thus, form a family. The distances detailed hereafter are defined over two vectors $P$, $Q$ over $\mathbb{R}^d$.
\begin{enumerate}
\item \textbf{$L_2$ Distance - Euclidean Distance.}
The Euclidean Distance is one of the most common distance measures used for numerical attributes or features in the Euclidean space. It derives from the Pythagorean theorem. It is a formal metric, defined by~\cite{deza2009encyclopedia}:
$$d_{L_2}(P,Q) = \sqrt{\sum^d_{i=1} |P_i -Q_i|^2}$$
For example: $d_{L_2}((5,3,4),(2,5,7))=\sqrt{(5-2)^2 + (3-5)^2 + (4-7)^2}=\sqrt{22}= 4.69$

\item \textbf{Squared Euclidean Distance.}\footnote{This distance was considered by~\cite{c:2007} as part of the $\chi^2$ distance family. We find the connection to the Minkowski family stronger, since it is based on the Euclidean Distance.}
In many applications, when considering the popular Euclidean Distance, it may be more convenient to omit the final square root in the calculation. The resulting value is called the Squared Euclidean Distance and is defined by \cite{deza2009encyclopedia}:
$$d_{L_2^2}(P,Q)= \sum^d_{i=1} (P_i - Q_i)^2$$
For example: $d_{L_2^2}((5,3,4),(2,5,7)) = 22$

\item \textbf{$L_1$ Distance - Manhattan Distance.}\footnote{Also known as Boxcar distance, City Block distance, Rectilinear distance and Absolute Value distance.}
Minkowski considered a non Euclidean distance, which cannot be calculated according to the Pythagorean theorem. An example of such a case is the distance between two locations in a city, which is the reason this metric is known as City Block Distance~\cite{k:1986}. Formally, 
$$d_{L_1}(P,Q) = \sum^d_{i=1} |P_i - Q_i|$$
For example: $d_{L_1}( (5,3,4),(2,5,7))$ $= 8$

The $L_1$ distance is not normalized, thus it can increase with the number of characteristics -- the vectors dimensions. It has many variants summing the absolute difference between the corresponding elements of the vectors, yet normalizing differently:
\begin{enumerate}
      \item \textbf{Gower Distance.}
      The Gower distance \cite{Gower:1971} seeks the average absolute difference between the corresponding elements of the vectors $P,Q$, while normalizing each absolute difference by the size of the range of the $i$th elements of the vectors, where the range is denoted by $R_i$ and its size by $|R_i|$. Formally,
      $$d_{Gow}(P,Q) = \frac{1}{d}\sum^d_{i=1}\frac{|P_i - Q_i|}{|R_i|}$$
     
     % For example, let $P = (5,3,4)$ and $Q = (2,5,7)$, where $R_1 = [2,6]$ and $|R_1|=5$, $R_2 = [1,6]$ and $|R_2|=6$, $R_3 = [2,10]$ and $|R_3|=9$, then,
      For example, let $P = (5,3,4)$ and $Q = (2,5,7)$, where $R_1 = [2,6]$, $R_2 = [1,6]$ and $R_3 = [2,10]$, then, $|R_1|=5, |R_2|= 6$ and $|R_3|=9$, and $d_{Gow}(P,Q) = \frac{1}{3 }(\frac{3}{5} + \frac{2}{6} + \frac{3}{9}) = \frac{19}{45} = 0.422$
 
      \item \textbf{Soergel Distance.}\footnote{Also known as the Ruzicka distance.}\label{SoergelRuzicka}
      The Soergel distance~\cite{wbSoergel:1998} normalizes the sum of the absolute differences between the corresponding elements of the vectors $P,Q$, by the sum of the maximal values of the corresponding elements of the vectors.
      Formally,
      $$d_{Soer}(P,Q) = \frac{\sum^d_{i=1} |P_i - Q_i|}{\sum^d_{i=1}\max (P_i,Q_i)}$$
      
      For example: $d_{Soer}((5,3,4), (2,5,7)) = \frac{3 + 2 + 3}{5 + 5 + 7} = \frac{8}{17} = 0.47$

      Soergel distance is identical to Jaccard/Tanimoto coefficient for binary variables.
    
      \item \textbf{Kulczynski Distance.}
      The Kulczynski distance~\cite{wbSoergel:1998} normalizes the sum of the absolute differences between the corresponding elements of the vectors $P,Q$,  by the sum of the minimal values of the corresponding elements of the vectors. Formally, 
      $$d_{Kul}(P,Q) = \frac{\sum^d_{i=1} |P_i - Q_i|}{\sum^d_{i=1}\min (P_i,Q_i)}$$
      For example: $d_{Kul}((5,3,4),(2,5,7)) = \frac{3 + 2 + 3}{2 + 3 + 4} = \frac{8}{9} = 0.889$

      \item \textbf{Canberra Distance.}
      The Canberra distance~\cite{lwCanberra:66}, %\cite{lwCanberra:67} 
      (called  after the capital of its introducers, in retrospect to the Manhattan distance), normalizes the  absolute difference between the corresponding elements of the vectors $P,Q$ by the sum of the absolute variable values prior to summing. It is sensitive to small changes near zero. Formally, 
      $$d_{Can}(P,Q) =\sum^d_{i=1} \frac{ |P_i - Q_i|}{|P_i| +| Q_i|}$$
      
      For example: $d_{Can}((5,3,4),(2,5,7)) = \frac{3 }{7} + \frac{2}{8} + \frac{3}{11}= \frac{293}{308} = 0.951$

      The Canberra distance has a variant called Adkins form, in which the distance is divided by $(n-Z)$, where $Z$ is the number of attributes that are 0 for $P$ and $Q$.

    \item \textbf{Lorentzian Distance.}\footnote{An alternative definition for this distance~\cite{Deza:2006} where the Lorentzian Distance is $d_{Lor} =\sum^d_{i=1} \ln ( 1 + | P_i - Q_i | )$.}
    The Lorentzian distance~\cite{Lorentzian:2016} gives a chance to change the contribution of the features. Different features may affect the distance in different proportions. This flexibility may give better results for classification problems. It is a special case, which may give a zero distance for points that are far in Euclidean Distance. Therefore, it is not a formal metric\footnote{Since the Lorentzian distance violates the positive definiteness property,~\cite{Lorentzian:2016} first perform pre-processing on the data-set points before computing the distance.}. Formally, 
    $$\sqrt{(\sum^{d-1}_{i=1} |P_i -Q_i|^2)-|P_d -Q_d|^2}$$
    For example: $d_{Lor}((5,3,4),(2,5,7)) = \sqrt{|5-2|^2 + |3-5|^2-|4-7|^2} = \sqrt{13-9} = 2$
\end{enumerate}

\item \textbf{$L_p$ Distance - Minkowski Distance.}
Minkowski generalized the Euclidean Distance for higher order exponents and roots up to a constant $p$, also known as $L_p$. Formally~\cite{deza2009encyclopedia}, 
$$d_{L_p}(P,Q) = \sqrt[p]{\sum^d_{i=1} |P_i -Q_i|^p}$$

For example, let $p=3$, then:
$$d_{L_3}((5,3,4),(2,5,7))= \sqrt[3]{(5-2)^3 + (3-5)^3 + (4-7)^3} = \sqrt[3]{62} = 3.958$$
Minkowski with $p=1$ and $p=2$ give Manhattan and Euclidean distances, respectively.

\item \textbf{$L_\infty$ Distance - Chebyshev Distance.}\footnote{Also known as lattice metric, Chessboard distance or the Minmax approximation.}
This is the $L_p$ distance, when $p$ goes to infinity. It is formally defined by \cite{deza2009encyclopedia}: 
$$d_{L_\infty}(P,Q) = \max_i |P_i - Q_i|$$

For example:  $d_{L_\infty}((5,3,4),(2,5,7)) = \max \{ 3, 2, 3\} = 3$
\end{enumerate}

\subsection{Intersection Similarity Measures and Distances}\label{ss:intersection}
In this subsection we review measures based on sets intersection. A basic notion is the probability density function, (\emph{PDF}), describing how likely a variable is to have a certain value. 

\begin{definition}\label{d:pdf} {\rm {\bf [Histogram and Probability Density Function (\emph{PDF})]}~\cite{2011iv}}
A \emph{histogram} counts the number of occurrences of a value in a given range. A \emph{PDF} is a histogram which is normalized by dividing each value by the total number of observations. A \emph{PDF} values sum is 1.	
\end{definition}

For the rest of this section, let $P$ and $Q$ be \emph{PDF}s. Most similarity measures pertinent to the intersection can be transformed into distance measures using the formula $d_x(P,Q) = 1 - s_x(P,Q)$.
\begin{enumerate}
\item \textbf{Intersection Similarity and Distance.}
Developing visual skills for robots allowing them to interact with a dynamic, realistic environment, required new kinds of vision algorithms. Given a discrete color space (e.g.,\ red, green, blue), the color histogram is obtained by discretizing the image colors and counting the number of times each discrete color occurs in the image array. Swain and Ballard~\cite{Swain:1991} introduced a method of comparing image and model histograms called \textbf{Histogram Intersection}, which describes how many of the pixels in the model histogram are found in the image. Formally, the intersection similarity between two probability density functions, $P$ and $Q$, is defined~\cite{Swain:1991,Cha-Srihari:2002} by: 
$$sim_{IS}(P,Q)=\sum^d_{i=1}\min(P_i,Q_i)$$ 

When scaling the image histogram to be the same size as the model histogram is possible, then Histogram Intersection is equivalent to the use of the Manhattan distance. That is, if $\sum^d_{i=1}P_i=\sum^d_{i=1}Q_i$ then we can define: $d_{IS}(P,Q)=1-sim_{IS}(P,Q)=\frac{1}{2}\sum^d_{i=1}|P_i-Q_i|$.

For example, let $P=(\frac{2}{14},\frac{3}{14},\frac{4}{14},\frac{5}{14})$ and $Q=(\frac{1}{14},\frac{2}{14},\frac{5}{14},\frac{6}{14})$, then,
$$sim_{IS}(P,Q) = \frac{1}{14}+\frac{2}{14}+\frac{4}{14}+\frac{5}{14}=\frac{12}{14}$$
and $$d_{IS}(P,Q)=1-sim_{IS}(P,Q)=\frac{2}{14}$$
Aside from the original two image similarity mentioned above, it has applications in feature selection, image indexing and retrieval, pattern classification and clustering.

\item \textbf{Wave Hedges Distance.}
The meaning of the distance formula is the sum of the differences between the total, that is 1, and the ratio between the common part (their minimum) of the histograms/\emph{PDF}s and their union (their maximum). Formally\footnote{Hassanat~\cite{Hassanat:2014} surveys several distance functions, including Wave-Hedges distance. Interestingly, he reports that the source of the Wave-Hedges metric has not been correctly cited. Some allude to it incorrectly as~\cite {Hedges:1976}, however, the source of this metric eludes Hassanat. Even the name of this measure is questioned. Nevertheless, since this distance is listed in many surveys, we chose to include it in this guide as well.}, 
$$d_{WH}(P,Q)=\sum^d_{i=1}(1-\frac{min(P_i,Q_i)}{max(P_i,Q_i)})$$ 
An alternative form based on a sum of absolute differences is defined by:
$$d_{WH}(P,Q)=\sum^d_{i=1}\frac{|P_i-Q_i|}{max(P_i,Q_i)}$$

For example, using the above $P$,$Q$, we get:
\commentout{
\begin{eqnarray}
\nonumber 
  d_{WH}(P,Q)&=&(1-\frac{\frac{1}{14}}{\frac{2}{14}})+(1-\frac{\frac{2}{14}}{\frac{3}{14}})+(1-\frac{\frac{4}{14}}{\frac{5}{14}})+(1-\frac{\frac{5}{14}}{\frac{6}{14}})\\
\nonumber 
  &=& \frac{1}{2}+\frac{1}{3}+\frac{1}{5}+\frac{1}{6}=\frac{15+10+6+5}{30}=1\frac{1}{5}
\end{eqnarray}
}
\begin{eqnarray*}
\nonumber 
d_{WH}(P,Q)&=&(1-\frac{\frac{1}{14}}{\frac{2}{14}})+(1-\frac{\frac{2}{14}}{\frac{3}{14}})+(1-\frac{\frac{4}{14}}{\frac{5}{14}})+(1-\frac{\frac{5}{14}}{\frac{6}{14}})=\frac{15+10+6+5}{30}\\
&=&1.2
\end{eqnarray*}
\item \textbf{S{\o}rensen Distance.}\footnote{Also known as Bray-Curtis distance or Czekanowski Coefficient.}
The Danish botanist and evolutionary biologist, Thorvald S{\o}rensen, was interested in a formal scientific way of measuring the similarity between groups of species in different vegetation environments. Let $P$ denote the distribution of a group of species in one environment, and $Q$ denote the distribution of the same group of species in another environment. Then, the similarity between the environments is defined by~\cite{sorensen:1948}:
$$sim_{Sor}(P,Q)=\frac{2\sum^d_{i=1}\min(P_i,Q_i)}{\sum^d_{i=1}(P_i+Q_i)}$$
The meaning of the formula is the ratio of the differences between the values in the co-related \emph{PDF}-bins and the \emph{PDF}s average (hence, the 2 in the numerator). It is widely used in ecology~\cite{Looman:1960}. The distance is defined by:
$$d_{Sor}(P,Q)=1-sim_{Sor}(P,Q)=\frac{\sum^d_{i=1}|P_i-Q_i|}{\sum^d_{i=1}(P_i+Q_i)}$$

For example, using the above $P$,$Q$, we get: $sim_{Sor}(P,Q)=\frac{2(\frac{1}{14}+\frac{2}{14}+\frac{4}{14}+\frac{5}{14})}{2}=\frac{2\cdot\frac{12}{14}}{2}=\frac{12}{14}$ and 
$d_{Sor}(P,Q)=1-sim_{Sor}(P,Q)= 1-\frac{12}{14}=\frac{1}{7}$.

The following is a variant of the Sørensen distance:
\begin{enumerate}
\commentout{
\item \textbf{Czekanowski Coefficient}
In 1909, Czekanowski aimed at separating anthropological finding into groups of populations that stand out from one another. Some articles identify his similarity measure with S{\o}rensen's, even though they are not identical in all cases.

Czekanowski Coefficient is defined~\cite{Czekanowski:1909} by $$sim_{Cze}=\frac{2\sum^d_{i=1}min(P_i,Q_i)}{\sum^d_{i=1}(P_i+Q_i)}$$ 
The intersection meaning of the similarity calculated is the ratio between the common part of the histograms/\emph{PDF}s and their average (hence, the 2 in the numerator).
Its distance form is identical to Sørensen and is defined by: $$d_{Cze}(P,Q)=1-sim_{Cze}(P,Q)=\frac{\sum^d_{i=1}|P_i-Q_i|}{\sum^d_{i=1}(P_i+Q_i)}$$

For example, let $P=(\frac{2}{14},\frac{3}{14},\frac{4}{14},\frac{5}{14})$ and $Q=(\frac{1}{14},\frac{2}{14},\frac{5}{14},\frac{6}{14})$, then,
$$sim_{Cze}(P,Q)=\frac{1+2+4+5}{14}=\frac{12}{14}=\frac{6}{7}$$ 
and 
$$d_{Cze}(P,Q)=1-sim_{Cze}(P,Q)= \frac{1}{7}$$
}
\item \textbf{Motyka Similarity.}
It is half of the Sørensen distance (Czekanowski Coefficient). Formally~\cite{Motyka:1950}, $$sim_{Mot}(P,Q)=\frac{\sum^d_{i=1}min(P_i,Q_i)}{\sum^d_{i=1}(P_i+Q_i)}$$

Its distance form is given by: 
$$d_{Mot}(P,Q)=1-sim_{Mot}(P,Q)=\frac{\sum^d_{i=1}max(P_i,Q_i)}{\sum^d_{i=1}(P_i+Q_i)}$$
\commentout{
For example, let $P=(\frac{2}{14},\frac{3}{14},\frac{4}{14},\frac{5}{14})$ and $Q=(\frac{1}{14},\frac{2}{14},\frac{5}{14},\frac{6}{14})$, then,
$$sim_{Mot}(P,Q)=\frac{1+2+4+5}{14}/2=\frac{12}{14}/2=\frac{3}{7}$$ 
and 
$$d_{Mot}(P,Q)=1-sim_{Mot}(P,Q)=\frac{4}{7}$$}
\end{enumerate}

\item \textbf{Kulczynski Similarity.}\footnote{Also appears in Subsection~\ref{ss:Minkowski} for vectors. Here it is defined on \emph{PDF}s.}
Kulczynski wanted to estimate floristic similarity of plant sociology. Formally~\cite{kulczynski:1927}, 
$$sim_{Kul}(P,Q)=\frac{\sum^d_{i=1}min(P_i,Q_i)}{\sum^d_{i=1}|P_i-Q_i|}$$
The meaning of the formula is the ratio between the sum of the common parts (the minimum) and the sum of the absolute differences. Unlike the other similarity-distance relationships, Kulczynski similarity-distance relationship is defined by: 
$$sim_{kul}(P,Q) = \frac{1}{d_{Kul}(P,Q)}=\frac{\sum^d_{i=1}|P_i-Q_i|}{\sum^d_{i=1}min(P_i,Q_i)}$$

For example, using the above $P$,$Q$, we get: $sim_{Kul}(P,Q)=\frac{\frac{12}{14}}{\frac{4}{14}}=3$ and
$d_{Kul}(P,Q)=\frac{1}{sim_{Kul}(P,Q)}=\frac{1}{3}$.

\item \textbf{Jaccard (Tanimoto) Index.}\footnote{Also appears in Subsection~\ref{ss:Minkowski} for vectors. Here it is defined on \emph{PDF}s.}
Paul Jaccard~\cite{Jaccard}, %\cite{Jaccard:1901}
 suggested this measure for analysing the distribution of flora in the alpine zone. The same measure was independently formulated by Tanimoto\cite{tani:58} defining a classification and prediction tool. Formally,
\begin{eqnarray*}
\nonumber
d_{Jac}(P,Q) &=&\frac{\sum^d_{i=1}P_i+\sum^d_{i=1}Q_i-2\sum^d_{i=1}min(P_i,Q_i)}{\sum^d_{i=1}P_i+\sum^d_{i=1}Q_i-\sum^d_{i=1}min(P_i,Q_i)}\\
\nonumber
&=& \frac{\sum^d_{i=1}(max(P_i,Q_i)-min(P_i,Q_i))}{\sum^d_{i=1}max(P_i,Q_i)} 
\end{eqnarray*} 

The meaning of the formula is the ratio between the sum of differences of the values in the co-related \emph{PDF}-bins and the sum of the maximum values. Ruzicka\footnote{Also known as Soergel, appears in Subsection~\ref{SoergelRuzicka} for vectors.} defined a similarity measure which completes the Jaccard distance~\cite{Ruzicka:1958,Soergel:1967} by: 
$$sim_{Ruz}(P,Q)=1 - d_{Jac}(P,Q)=\frac{\sum^d_{i=1}min(P_i,Q_i)}{\sum^d_{i=1}max(P_i,Q_i)}$$

The meaning of the formula is the ratio between the sum of the common parts of the co-related \emph{PDF}-bins and the sum of maximum values.

For example, using the above $P$,$Q$, we get: $sim_{Ruz}(P,Q)=\frac{\frac{12}{14}}{\frac{16}{14}}=\frac{3}{4}$ and $d_{Jac}(P,Q)=1-sim_{Ruz}(P,Q)=1-\frac{3}{4}=\frac{1}{4}$
\end{enumerate}

\subsection{Entropy Family Measures}\label{ss:entropy}
In this subsection, we assume that $P$ and $Q$ are discrete probability distributions defined using a probability density function, \emph{PDF}, as defined in Subsection~\ref{ss:intersection}.
The measures in this family are based on the \emph{Shannon Entropy} (SE) defined on a given \emph{PDF} $P$ as follows:
$$SE(P)=\sum_{i=1}^{d}P_i\ln{P_i}$$
Entropy is a way to measure diversity. A diversity index is a quantitative statistical measure of how many different types exist in a data-set, accounting for community richness, evenness, and dominance. The Shannon entropy (index) is related to the proportional abundances of types. 

Entropy measure is also useful in machine learning. In general, entropy is a measure of uncertainty and the objective of machine learning is to minimize uncertainty. Decision tree learning algorithms use relative entropy (see below) to determine the decision rules that govern the data at each node.

The measures below make use of this concept to evaluate the difference between two given \emph{PDF}s.
\begin{enumerate}
\item \textbf{Kullback-Leibler Divergence -- Relative Entropy.}
The Kullback–Leibler divergence is a measure of how one probability distribution $Q$ is different from a second, reference probability distribution $P$. A simple interpretation of the divergence of $P$ from $Q$ is the expected excess surprise from using $Q$ as a model when the actual distribution is $P$. In the context of machine learning, the relative entropy is often called the \emph{information gain} achieved if $P$ would be used instead of $Q$ which is currently used.

Formally, let $P$ and $Q$ be two \emph{PDF}s, the Kullback–Leibler divergence is defined~\cite{KL:1951} by:
$$d_{\mathit{KL}}(P,Q)=\sum_{i=1}^{d}P_i\ln{\frac{P_i}{Q_i}}$$

Classification in machine learning performed by Logistic regression or Artificial neural networks often employs a standard loss function, called \emph{Cross entropy loss}, that minimizes the average cross entropy between ground truth and predicted distributions~\cite{RK:13}. In general, cross entropy is a measure of the differences between two data-sets similar to relative entropy. The \emph{cross entropy} (CE) of $P$ and $Q$ is defined by:
$$CE(P,Q)=-\sum_{i=1}^{d}P_i\ln{Q_i}=d_{\mathit{KL}}(P,Q)-SE(P)$$

For example, let $P=(\frac{2}{14},\frac{3}{14},\frac{4}{14},\frac{5}{14})$ and $Q=(\frac{1}{14},\frac{2}{14},\frac{5}{14},\frac{6}{14})$, then,
\commentout{
\begin{eqnarray*}
\nonumber
d_{\mathit{KL}}(P,Q)&=&\frac{2}{14}\ln{\frac{\frac{2}{14}}{\frac{1}{14}}}+\frac{3}{14}\ln{\frac{\frac{3}{14}}{\frac{2}{14}}}+\frac{4}{14}\ln{\frac{\frac{4}{14}}{\frac{5}{14}}}+\frac{5}{14}\ln{\frac{\frac{5}{14}}{\frac{6}{14}}}\\
\nonumber
 &=&\frac{2}{14}\ln{2}+\frac{3}{14}\ln{\frac{3}{2}}+\frac{4}{14}\ln{\frac{4}{5}}+\frac{5}{14}\ln{\frac{5}{6}}\\
\nonumber
 &=& 0.099+0.087+(-0.064)+(-0.065)=0.057
\end{eqnarray*}
}
\begin{eqnarray*}
\nonumber
d_{\mathit{KL}}(P,Q)&=&\frac{2}{14}\ln{\frac{\frac{2}{14}}{\frac{1}{14}}}+\frac{3}{14}\ln{\frac{\frac{3}{14}}{\frac{2}{14}}}+\frac{4}{14}\ln{\frac{\frac{4}{14}}{\frac{5}{14}}}+\frac{5}{14}\ln{\frac{\frac{5}{14}}{\frac{6}{14}}}=0.057
\end{eqnarray*}

It is not a metric: it is asymmetric and does not satisfy the triangle inequality.

\item \textbf{Jeffreys-Divergence (J-Divergence).}
Jeffrey's divergence is a symmetric version of the Kullback–Leibler divergence, since it is equal to $d_{\mathit{KL}}(P,Q)+d_{\mathit{KL}}(Q,P)$. It is used in for wide applications, from change detection to clutter homogeneity analysis in radar processing.

Formally, the J-Divergence is defined~\cite{JEF:1948} by:
$$d_{\mathit{J}}(P,Q)=\sum_{i=1}^{d}P_i\ln{\frac{P_i}{Q_i}}+\sum_{i=1}^{d}Q_i\ln{\frac{Q_i}{P_i}}=\sum_{i=1}^{d}(P_i-Q_i)\ln{\frac{P_i}{Q_i}}$$

For example, using the above $P$,$Q$, we get: $d_{\mathit{J}}(P,Q)=$
\commentout{
\begin{eqnarray*}
\nonumber
d_{\mathit{J}}(P,Q)&=&(\frac{2}{14}-\frac{1}{14})\ln{\frac{\frac{2}{14}}{\frac{1}{14}}}+(\frac{3}{14}-\frac{2}{14})\ln{\frac{\frac{3}{14}}{\frac{2}{14}}}+(\frac{4}{14}-\frac{5}{14})\ln{\frac{\frac{4}{14}}{\frac{5}{14}}}+(\frac{5}{14}-\frac{6}{14})\ln{\frac{\frac{5}{14}}{\frac{6}{14}}}\\
\nonumber
 &=&\frac{1}{14}\ln{2}+\frac{1}{14}\ln{\frac{3}{2}}+\frac{-1}{14}\ln{\frac{4}{5}}+\frac{-1}{14}\ln{\frac{5}{6}}\\
\nonumber
 &=& 0.0495+0.0289+0.016+0.013=0.1074
\end{eqnarray*}
}
\begin{eqnarray*}
\nonumber
&=&(\frac{2}{14}-\frac{1}{14})\ln{\frac{\frac{2}{14}}{\frac{1}{14}}}+(\frac{3}{14}-\frac{2}{14})\ln{\frac{\frac{3}{14}}{\frac{2}{14}}}+(\frac{4}{14}-\frac{5}{14})\ln{\frac{\frac{4}{14}}{\frac{5}{14}}}+(\frac{5}{14}-\frac{6}{14})\ln{\frac{\frac{5}{14}}{\frac{6}{14}}}\\
\nonumber
 &=&\frac{1}{14}\ln{2}+\frac{1}{14}\ln{\frac{3}{2}}+\frac{-1}{14}\ln{\frac{4}{5}}+\frac{-1}{14}\ln{\frac{5}{6}}=0.1074
\end{eqnarray*}

\item \textbf{K-Divergence.}
The K-divergence measure is based on the Kullback–Leibler divergence, however, it refers to the mean of the two distributions instead of just the second. It actually equals the following formula: $d_{\mathit{KL}}(P,\frac{P+Q}{2})$. Formally, the K-divergence is defined~\cite{c:2007} by:
$$d_{\mathit{Kdiv}}(P,Q)=\sum_{i=1}^{d}P_i\ln{\frac{P_i}{\frac{P_i+Q_i}{2}}}=\sum_{i=1}^{d}P_i\ln{\frac{2P_i}{P_i+Q_i}}$$

For example, using the above $P$,$Q$, we get:
\commentout{
\begin{eqnarray*}
\nonumber
d_{\mathit{Kdiv}}(P,Q)&=&\frac{2}{14}\ln{\frac{2\cdot\frac{2}{14}}{\frac{2}{14}+\frac{1}{14}}}+\frac{3}{14}\ln{\frac{2\cdot\frac{3}{14}}{\frac{3}{14}+\frac{2}{14}}}+\frac{4}{14}\ln{\frac{2\cdot\frac{4}{14}}{\frac{4}{14}+\frac{5}{14}}}+\frac{5}{14}\ln{\frac{2\cdot\frac{5}{14}}{\frac{5}{14}+\frac{6}{14}}}\\
\nonumber
 &=&\frac{2}{14}\ln{\frac{4}{3}}+\frac{3}{14}\ln{\frac{6}{5}}+\frac{4}{14}\ln{\frac{8}{9}}+\frac{5}{14}\ln{\frac{10}{11}}\\
\nonumber
 &=& 0.041+0.039+(-0.034)+(-0.034)=0.012
\end{eqnarray*}
}
\begin{eqnarray*}
\nonumber
d_{\mathit{Kdiv}}(P,Q)&=&\frac{2}{14}\ln{\frac{2\cdot\frac{2}{14}}{\frac{2}{14}+\frac{1}{14}}}+\frac{3}{14}\ln{\frac{2\cdot\frac{3}{14}}{\frac{3}{14}+\frac{2}{14}}}+\frac{4}{14}\ln{\frac{2\cdot\frac{4}{14}}{\frac{4}{14}+\frac{5}{14}}}+\frac{5}{14}\ln{\frac{2\cdot\frac{5}{14}}{\frac{5}{14}+\frac{6}{14}}}\\
&=&0.012
\end{eqnarray*}
\item \textbf{Tops{\o}e Divergence.}\footnote{Also called Information statistics.}
Tops{\o}e divergence is similar to the K-divergence measure, however, it is symmetric in $P$ and $Q$. Formally, the Tops{\o}e Divergence is defined as follows~\cite{Top:2000}:
$$d_{\mathit{Top}}(P,Q)=\sum_{i=1}^{d}(P_i\ln{\frac{2P_i}{P_i+Q_i}}+Q_i\ln{\frac{2Q_i}{P_i+Q_i}})$$

For example, using the above $P$,$Q$, we get:
\commentout{
\begin{eqnarray*}
\nonumber
d_{\mathit{Top}}(P,Q)&=&\frac{2}{14}\ln{\frac{2\cdot\frac{2}{14}}{\frac{2}{14}+\frac{1}{14}}}+\frac{1}{14}\ln{\frac{2\cdot\frac{1}{14}}{\frac{2}{14}+\frac{1}{14}}}\\
\nonumber
&+&\frac{3}{14}\ln{\frac{2\cdot\frac{3}{14}}{\frac{3}{14}+\frac{2}{14}}}+\frac{2}{14}\ln{\frac{2\cdot\frac{2}{14}}{\frac{3}{14}+\frac{2}{14}}}\\
\nonumber
&+&\frac{4}{14}\ln{\frac{2\cdot\frac{4}{14}}{\frac{4}{14}+\frac{5}{14}}}+\frac{5}{14}\ln{\frac{2\cdot\frac{5}{14}}{\frac{4}{14}+\frac{5}{14}}}\\
\nonumber
&+&\frac{5}{14}\ln{\frac{2\cdot\frac{5}{14}}{\frac{5}{14}+\frac{6}{14}}}+\frac{6}{14}\ln{\frac{2\cdot\frac{6}{14}}{\frac{5}{14}+\frac{6}{14}}}\\
\nonumber
 &=&\frac{2}{14}\ln{\frac{4}{3}}+\frac{1}{14}\ln{\frac{2}{3}}+\frac{3}{14}\ln{\frac{6}{5}}+\frac{2}{14}\ln{\frac{4}{5}}+\frac{4}{14}\ln{\frac{8}{9}}\\
\nonumber
&+&\frac{5}{14}\ln{\frac{10}{9}}+\frac{5}{14}\ln{\frac{10}{11}}+\frac{6}{14}\ln{\frac{12}{11}}\\
\nonumber
 &=& 0.041-0.029+0.039-0.032-0.034+0.038-0.034+0.037\\
\nonumber
&=&0.026
\end{eqnarray*}
}
\begin{eqnarray*}
\nonumber
d_{\mathit{Top}}(P,Q)&=&\frac{2}{14}\ln{\frac{2\cdot\frac{2}{14}}{\frac{2}{14}+\frac{1}{14}}}+\frac{1}{14}\ln{\frac{2\cdot\frac{1}{14}}{\frac{2}{14}+\frac{1}{14}}}+\frac{3}{14}\ln{\frac{2\cdot\frac{3}{14}}{\frac{3}{14}+\frac{2}{14}}}+\frac{2}{14}\ln{\frac{2\cdot\frac{2}{14}}{\frac{3}{14}+\frac{2}{14}}}\\
\nonumber
&+&\frac{4}{14}\ln{\frac{2\cdot\frac{4}{14}}{\frac{4}{14}+\frac{5}{14}}}+\frac{5}{14}\ln{\frac{2\cdot\frac{5}{14}}{\frac{4}{14}+\frac{5}{14}}}+\frac{5}{14}\ln{\frac{2\cdot\frac{5}{14}}{\frac{5}{14}+\frac{6}{14}}}+\frac{6}{14}\ln{\frac{2\cdot\frac{6}{14}}{\frac{5}{14}+\frac{6}{14}}}\\
&=&0.026
\end{eqnarray*}
\item \textbf{Jensen-Shannon Divergence.}\footnote{Also known as Information Radius (IRad) or Total Divergence to the Average.}
The Jensen–Shannon divergence is another method of measuring the similarity between two probability distributions by averaging the measures of how different each distribution is from their average distribution. It is based on the Kullback–Leibler divergence, with useful differences, including that it is symmetric and it always has a finite value, and equals: $\frac{1}{2}d_{\mathit{KL}}(P,\frac{P+Q}{2})+\frac{1}{2}d_{\mathit{KL}}(Q,\frac{P+Q}{2})$. The square root of the Jensen–Shannon divergence is a metric often referred to as Jensen-Shannon distance. Formally, the Jensen–Shannon divergence is defined as follows~\cite{Lin:1991}:
$$d_{\mathit{JS}}(P,Q)=\frac{1}{2}[\sum_{i=1}^{d}P_i\ln{\frac{2P_i}{P_i+Q_i}}+\sum_{i=1}^{d}Q_i\ln{\frac{2Q_i}{P_i+Q_i}}]$$

For example, using the above $P$,$Q$, we get:
\commentout{
\begin{eqnarray*}
\nonumber
d_{\mathit{JS}}(P,Q)&=&\frac{1}{2}[\frac{2}{14}\ln{\frac{2\cdot\frac{2}{14}}{\frac{2}{14}+\frac{1}{14}}}+\frac{3}{14}\ln{\frac{2\cdot\frac{3}{14}}{\frac{3}{14}+\frac{2}{14}}}+\frac{4}{14}\ln{\frac{2\cdot\frac{4}{14}}{\frac{4}{14}+\frac{5}{14}}}+\frac{5}{14}\ln{\frac{2\cdot\frac{5}{14}}{\frac{5}{14}+\frac{6}{14}}}\\
\nonumber
&+&\frac{1}{14}\ln{\frac{2\cdot\frac{1}{14}}{\frac{2}{14}+\frac{1}{14}}}+\frac{2}{14}\ln{\frac{2\cdot\frac{2}{14}}{\frac{3}{14}+\frac{2}{14}}}+\frac{5}{14}\ln{\frac{2\cdot\frac{5}{14}}{\frac{4}{14}+\frac{5}{14}}}+\frac{6}{14}\ln{\frac{2\cdot\frac{6}{14}}{\frac{5}{14}+\frac{6}{14}}}]\\
\nonumber
 &=&\frac{1}{2}[\frac{2}{14}\ln{\frac{4}{3}}+\frac{3}{14}\ln{\frac{6}{5}}+\frac{4}{14}\ln{\frac{8}{9}}+\frac{5}{14}\ln{\frac{10}{11}}\\
\nonumber
&+& \frac{1}{14}\ln{\frac{2}{3}}+\frac{2}{14}\ln{\frac{4}{5}}+\frac{5}{14}\ln{\frac{10}{9}}+\frac{6}{14}\ln{\frac{12}{11}}]\\
\nonumber
 &=&\frac{1}{2}[0.041+0.039-0.034-0.034-0.029-0.031+0.038+0.037]\\
\nonumber
 &=&\frac{1}{2}\cdot 0.061=0.0305
\end{eqnarray*}
}
\begin{eqnarray*}
\nonumber
d_{\mathit{JS}}(P,Q)&=&\frac{1}{2}[\frac{2}{14}\ln{\frac{2\cdot\frac{2}{14}}{\frac{2}{14}+\frac{1}{14}}}+\frac{3}{14}\ln{\frac{2\cdot\frac{3}{14}}{\frac{3}{14}+\frac{2}{14}}}+\frac{4}{14}\ln{\frac{2\cdot\frac{4}{14}}{\frac{4}{14}+\frac{5}{14}}}+\frac{5}{14}\ln{\frac{2\cdot\frac{5}{14}}{\frac{5}{14}+\frac{6}{14}}}\\
\nonumber
&+&\frac{1}{14}\ln{\frac{2\cdot\frac{1}{14}}{\frac{2}{14}+\frac{1}{14}}}+\frac{2}{14}\ln{\frac{2\cdot\frac{2}{14}}{\frac{3}{14}+\frac{2}{14}}}+\frac{5}{14}\ln{\frac{2\cdot\frac{5}{14}}{\frac{4}{14}+\frac{5}{14}}}+\frac{6}{14}\ln{\frac{2\cdot\frac{6}{14}}{\frac{5}{14}+\frac{6}{14}}}]\\
&=&0.0305
\end{eqnarray*}

\item \textbf{Jensen Difference.}
The Jensen Difference method is to measure the sum of differences between the average of the information in the two given distributions and the information of their average. Formally, the Jensen Difference is defined as follows~\cite{TANEJA}:
$$d_{\mathit{Jdiff}}(P,Q)=\sum_{i=1}^{d}[\frac{P_i\ln{P_i}+Q_i\ln{Q_i}}{2}-(\frac{P_i+Q_i}{2})\ln{(\frac{P_i+Q_i}{2})}]$$

For example, using the above $P$,$Q$, we get:
\commentout{
\begin{eqnarray*}
\nonumber
d_{\mathit{Jdiff}}(P,Q)&=&\frac{\frac{2}{14}\ln{\frac{2}{14}}+\frac{1}{14}\ln{\frac{1}{14}}}{2}-\frac{\frac{2}{14}+\frac{1}{14}}{2}\ln{\frac{\frac{2}{14}+\frac{1}{14}}{2}}\\
\nonumber
&+&\frac{\frac{3}{14}\ln{\frac{3}{14}}+\frac{2}{14}\ln{\frac{2}{14}}}{2}-\frac{\frac{3}{14}+\frac{2}{14}}{2}\ln{\frac{\frac{3}{14}+\frac{2}{14}}{2}}\\
\nonumber
&+&\frac{\frac{4}{14}\ln{\frac{4}{14}}+\frac{5}{14}\ln{\frac{5}{14}}}{2}-\frac{\frac{4}{14}+\frac{5}{14}}{2}\ln{\frac{\frac{4}{14}+\frac{5}{14}}{2}}\\
\nonumber
&+&\frac{\frac{5}{14}\ln{\frac{5}{14}}+\frac{6}{14}\ln{\frac{6}{14}}}{2}-\frac{\frac{5}{14}+\frac{6}{14}}{2}\ln{\frac{\frac{5}{14}+\frac{6}{14}}{2}}\\
\nonumber
 &=&\frac{-0.278-0.189}{2}-0.107\ln{0.107}+\frac{-0.33-0.278}{2}-0.179\ln{0.179}\\
\nonumber
&+&\frac{-0.358-0.368}{2}-0.321\ln{0.321}+\frac{-0.368-0.363}{2}-0.393\ln{0.393}\\
\nonumber
 &=& -0.2335-(-0.239)-0.304-(-0.308)\\
\nonumber
&+&-0.363-(-0.365)-0.3655-(-0.367)=0.013
\end{eqnarray*}
}
\begin{eqnarray*}
\nonumber
d_{\mathit{Jdiff}}(P,Q)&=&\frac{\frac{2}{14}\ln{\frac{2}{14}}+\frac{1}{14}\ln{\frac{1}{14}}}{2}-\frac{\frac{2}{14}+\frac{1}{14}}{2}\ln{\frac{\frac{2}{14}+\frac{1}{14}}{2}}\\
\nonumber
&+&\frac{\frac{3}{14}\ln{\frac{3}{14}}+\frac{2}{14}\ln{\frac{2}{14}}}{2}-\frac{\frac{3}{14}+\frac{2}{14}}{2}\ln{\frac{\frac{3}{14}+\frac{2}{14}}{2}}\\
\nonumber
&+&\frac{\frac{4}{14}\ln{\frac{4}{14}}+\frac{5}{14}\ln{\frac{5}{14}}}{2}-\frac{\frac{4}{14}+\frac{5}{14}}{2}\ln{\frac{\frac{4}{14}+\frac{5}{14}}{2}}\\
\nonumber
&+&\frac{\frac{5}{14}\ln{\frac{5}{14}}+\frac{6}{14}\ln{\frac{6}{14}}}{2}-\frac{\frac{5}{14}+\frac{6}{14}}{2}\ln{\frac{\frac{5}{14}+\frac{6}{14}}{2}}=0.013
\end{eqnarray*}

\item \textbf{Structural Entropic Distance (SED).}
The Structural Entropic Distance (SED) was developed as a distance metric of an unordered tree structures~\cite{ConnorSI09}, and has been extended to handle other structured data as well. It is a ratio of the complexity of the mean vector to the geometric mean of complexities of the individual vectors, where the complexity is defined in terms of Shannon entropy -- the amount of information needed to describe the vector. Formally, the Structural Entropic Distance (SED) is defined as follows~\cite{ConnorSI09}:
$$d_{\mathit{SED}}(P,Q)=\frac{C(\frac{P+Q}{2})}{\sqrt{C(P)C(Q)}}-1,$$
where $C(X)=b^{\sum_iX_i\log_bX_i}$ when $b$ is any constant base. Note that when $P=Q$ then $d_{\mathit{SED}}=0$.

For example, using the above $P$,$Q$, then taking $b$ to be the natural base we get,
\commentout{
\begin{eqnarray*}
\nonumber
C(\frac{P+Q}{2})&=&\exp^{\sum_i\frac{P_i+Q_i}{2}\ln{\frac{P_i+Q_i}{2}}}\\
\nonumber
&=&\exp^{\frac{\frac{2}{14}+\frac{1}{14}}{2}\ln{\frac{\frac{2}{14}+\frac{1}{14}}{2}}+\frac{\frac{3}{14}+\frac{2}{14}}{2}\ln{\frac{\frac{3}{14}+\frac{2}{14}}{2}}+\frac{\frac{4}{14}+\frac{5}{14}}{2}\ln{\frac{\frac{4}{14}+\frac{5}{14}}{2}}+\frac{\frac{5}{14}+\frac{6}{14}}{2}\ln{\frac{\frac{5}{14}+\frac{6}{14}}{2}}}\\
\nonumber
&=&\exp^{\frac{3}{28}\ln{\frac{3}{28}}+\frac{5}{28}\ln{\frac{5}{28}}+\frac{9}{28}\ln{\frac{9}{28}}+\frac{11}{28}\ln{\frac{11}{28}}}\\
\nonumber
 &=&\exp^{-0.239-0.3077-0.648-0.3671}=\exp^{-1.5617}=0.2098
\end{eqnarray*}
}
\begin{eqnarray*}
\nonumber
C(\frac{P+Q}{2})=\exp^{\sum_i\frac{P_i+Q_i}{2}\ln{\frac{P_i+Q_i}{2}}}=\exp^{\frac{3}{28}\ln{\frac{3}{28}}+\frac{5}{28}\ln{\frac{5}{28}}+\frac{9}{28}\ln{\frac{9}{28}}+\frac{11}{28}\ln{\frac{11}{28}}}=0.2098
\end{eqnarray*}
In addition,
\commentout{
\begin{eqnarray*}
\nonumber
C(P)&=&\exp^{\sum_i P_i\ln P_i}\\
\nonumber
&=&\exp^{\frac{2}{14}\ln{\frac{2}{14}}+\frac{3}{14}\ln{\frac{3}{14}}+\frac{4}{14}\ln{\frac{4}{14}}+\frac{5}{14}\ln{\frac{5}{14}}}\\
\nonumber
&=&\exp^{-0.278-0.330-0.358-0.368}=\exp^{-1.334}=0.2635
\end{eqnarray*}
}
\begin{eqnarray*}
\nonumber
C(P)&=&\exp^{\sum_i P_i\ln P_i}=\exp^{\frac{2}{14}\ln{\frac{2}{14}}+\frac{3}{14}\ln{\frac{3}{14}}+\frac{4}{14}\ln{\frac{4}{14}}+\frac{5}{14}\ln{\frac{5}{14}}}=0.2635
\end{eqnarray*}
and
\commentout{
\begin{eqnarray*}
\nonumber
C(Q)&=&\exp^{\sum_i Q_i\ln Q_i}\\
\nonumber
&=&\exp^{\frac{1}{14}\ln{\frac{1}{14}}+\frac{2}{14}\ln{\frac{2}{14}}+\frac{5}{14}\ln{\frac{5}{14}}+\frac{6}{14}\ln{\frac{6}{14}}}\\
\nonumber
&=&\exp^{-0.1885-0.278-0.368-0.363}=\exp^{-1.1976}=0.3019\\
\end{eqnarray*}
}
\begin{eqnarray*}
\nonumber
C(Q)&=&\exp^{\sum_i Q_i\ln Q_i}=\exp^{\frac{1}{14}\ln{\frac{1}{14}}+\frac{2}{14}\ln{\frac{2}{14}}+\frac{5}{14}\ln{\frac{5}{14}}+\frac{6}{14}\ln{\frac{6}{14}}}=0.3019
\end{eqnarray*}
Therefore,
$$d_{\mathit{SED}}(P,Q)=\frac{0.2098}{\sqrt{0.2635\cdot 0.3019}}-1=\frac{0.2098}{0.2821}-1=-0.2208$$
\end{enumerate}

\subsection{The $\chi ^2$ Family Measures}\label{ss:ChiSquared}
The chi-squared ($\chi^2$) test~\cite{ChiSquared:1996,ChiSquared:2013} is a statistical test used to compare the distributions of a categorical variable in two different samples. It originates in calculating the squared difference between the observed distribution and the expected distribution normalized by the expected one. Formally,
$$d_{\chi^2} =  \frac{(O - E)^2}{E}$$ where $O$ represents the observed distribution and $E$ represents the expected value.

The $\chi^2$ test is used in statistics for categorical variables in two ways: [a.] A test of independence: checking if the distribution of the categorical variable is not much different over distinct groups. If so, we can conclude that the categorical variable and groups are independent. For example, if more men than women have a specific condition, there is bigger chance to find a person with the condition among men than among women. In this case, we don't consider the gender to be independent of the condition. If there is equal chance of having the condition among men and women, we consider gender to be independent of the condition. [b.] A goodness of fit test: checking if  a variable  is likely to come from a specified distribution or not. It is often used to evaluate whether sample data is representative of the population.

In this family of measures, the input vectors are probability density functions (\emph{PDF}s).
\begin{enumerate}
\item \textbf{Pearson Distance and Correlation.}
As mentioned, Pearson's Distance~\cite{Pearson:1900} is the known $\chi^2$ test, where the second \emph{PDF} input is considered as the expected distribution used as a normalization for the squared distance. It can be interpreted as describing the range between no association (0 value) and a perfect monotonic relationship (–1/+1 values). Formally, 
$$d_{\mathit{Pear}}(P,Q) = \sum^d_{i=1}\frac{(P_i - Q_i)^2}{Q_i}$$

For example, let $P=(\frac{2}{14},\frac{3}{14},\frac{4}{14},\frac{5}{14})$ and $Q=(\frac{1}{14},\frac{2}{14},\frac{5}{14},\frac{6}{14})$, then,
\commentout{
$$d_{\mathit{Pear}}(P,Q) = \frac{(\frac{2}{14}- \frac{1}{14})^2 }{\frac{1}{14}} + \frac{(\frac{3}{14}- \frac{2}{14})^2 }{\frac{2}{14}} + \frac{(\frac{4}{14}- \frac{5}{14})^2 }{\frac{5}{14}} + \frac{(\frac{5}{14}- \frac{6}{14})^2 }{\frac{6}{14}}$$
$$=\frac{1}{14} + \frac{1}{28} + \frac{1}{70}+  \frac{1}{84}= \frac{28}{210} = 0.133$$
}
$$d_{\mathit{Pear}}(P,Q) = \frac{(\frac{2}{14}- \frac{1}{14})^2 }{\frac{1}{14}} + \frac{(\frac{3}{14}- \frac{2}{14})^2 }{\frac{2}{14}} + \frac{(\frac{4}{14}- \frac{5}{14})^2 }{\frac{5}{14}} + \frac{(\frac{5}{14}- \frac{6}{14})^2 }{\frac{6}{14}} = 0.133$$
In addition to Pearson's distance measure, the \textbf{Pearson's correlation coefficient}, denoted by $\rho$, is defined by the covariance of the two variables divided by the product of their standard deviations. It is a normalized covariance measure, having a value between $-1$ and $1$. Formally, let $\sigma_X$ is the standard deviation of a \emph{PDF} $X$, then $\rho(P,Q)$ is defined by:
$$\rho(P,Q)= \frac{\mathit{cov}(P,Q)}{\sigma_P\cdot \sigma_Q}$$
where $\mathit{cov}(P,Q)$ denotes the covariance of $P$ and $Q$, defined as follows. Let $E$ be the expectation and superfix $T$ denote matrix transpose operation, then: 
$$\mathit{cov}(P,Q)=E[(P-E[P])(Q-E[Q])^T]$$

The classical equivalent definition in probability theory for this measure is:
$$\rho(P,Q) = \frac{E[P\cdot Q] -  E[P]\cdot E[Q]}{\sqrt{E[P^2] -(E[P])^2}\cdot \sqrt{E[Q^2] -(E[Q])^2}}$$

As Pearson distance is asymmetric, some symmetric variations are detailed below.
\item \textbf{Neyman $\chi^2$.}
The Neyman $\chi^2$ Distance \cite{Neyman:1949} considers normalizing the squared difference between the given distributions pair by the first one. Formally,  
$$d_{\mathit{Ney}}(P,Q) = \sum^d_{i=1}\frac{(P_i - Q_i)^2}{P_i}$$

For example, using the above $P$,$Q$, we get:
\commentout{
$$d_{\mathit{Ney}}(P,Q) =\frac{(\frac{2}{14}- \frac{1}{14})^2 }{\frac{2}{14}} + \frac{(\frac{3}{14}- \frac{2}{14})^2 }{\frac{3}{14}} + \frac{(\frac{4}{14}- \frac{5}{14})^2 }{\frac{4}{14}} + \frac{(\frac{5}{14}- \frac{6}{14})^2 }{\frac{5}{14}}$$
$$ = \frac{1}{28} + \frac{1}{42} + \frac{1}{56} + \frac{1}{70} = \frac{77}{840}=0.092$$
}
$$d_{\mathit{Ney}}(P,Q) =\frac{(\frac{2}{14}- \frac{1}{14})^2 }{\frac{2}{14}} + \frac{(\frac{3}{14}- \frac{2}{14})^2 }{\frac{3}{14}} + \frac{(\frac{4}{14}- \frac{5}{14})^2 }{\frac{4}{14}} + \frac{(\frac{5}{14}- \frac{6}{14})^2 }{\frac{5}{14}}=0.092$$

\item \textbf{Additive Symmetric $\chi^2$.}
The Additive Symmetric $\chi^2$ \cite{jdivergance:2006} was formed by applying the J-Divergence (see Subsection~\ref{ss:entropy}), i.e.,\ considering the sum of two distances, applied to the same \emph{PDF} vectors in the $\chi^2$ distance, each in a different order. Formally,  
$$d_{AdSym\chi^2}(P,Q) = d_{Pear}(P,Q) + d_{Ney}(Q,P) = \sum^d_{i=1}\frac{(P_i - Q_i)^2(P_i + Q_i)}{P_iQ_i}$$

For example, using the above $P$,$Q$, we get:
\commentout{
$$d_{AdSym\chi^2}(P,Q) = \frac{1}{14^2}\cdot \frac{3}{14}\cdot \frac{14^2}{2} + 
                          \frac{1}{14^2}\cdot \frac{5}{14}\cdot \frac{14^2}{6} +
                          \frac{1}{14^2}\cdot \frac{9}{14}\cdot \frac{14^2}{20} +
                          \frac{1}{14^2}\cdot \frac{11}{14}\cdot \frac{14^2}{30}$$
                          $$=\frac{3}{28} + \frac{5}{84}+  \frac{9}{280} + \frac{11}{420} = \frac{63}{280}=0.225$$
}                          
\commentout{
$$d_{AdSym\chi^2}(P,Q) = \frac{(\frac{2}{14}-\frac{1}{14})^2(\frac{2}{14}+\frac{1}{14}) }{\frac{2}{14}\cdot \frac{1}{14}} + \frac{(\frac{3}{14}-\frac{2}{14})^2(\frac{3}{14}+\frac{2}{14}) }{\frac{3}{14}\cdot \frac{2}{14}} +$$
$$\frac{(\frac{4}{14}-\frac{5}{14})^2(\frac{4}{14}+\frac{5}{14}) }{\frac{4}{14}\cdot \frac{5}{14}} + \frac{(\frac{5}{14}-\frac{6}{14})^2(\frac{5}{14}+\frac{6}{14}) }{\frac{5}{14}\cdot \frac{6}{14}}$$
$$=\frac{\frac{1}{196} \cdot \frac{3}{14} }{\frac{2}{196}} + \frac{\frac{1}{196} \cdot \frac{5}{14} }{\frac{6}{196}} + \frac{\frac{1}{196} \cdot \frac{9}{14} }{\frac{20}{196}} +
  \frac{\frac{1}{196} \cdot \frac{11}{14} }{\frac{30}{196}}$$
$$=\frac{3}{28} + \frac{5}{84}+  \frac{9}{280} + \frac{11}{420} = \frac{63}{280}=0.225$$
}
$$d_{AdSym\chi^2}(P,Q) = \frac{(\frac{2}{14}-\frac{1}{14})^2(\frac{2}{14}+\frac{1}{14}) }{\frac{2}{14}\cdot \frac{1}{14}} + \frac{(\frac{3}{14}-\frac{2}{14})^2(\frac{3}{14}+\frac{2}{14}) }{\frac{3}{14}\cdot \frac{2}{14}} +$$
$$\frac{(\frac{4}{14}-\frac{5}{14})^2(\frac{4}{14}+\frac{5}{14}) }{\frac{4}{14}\cdot \frac{5}{14}} + \frac{(\frac{5}{14}-\frac{6}{14})^2(\frac{5}{14}+\frac{6}{14}) }{\frac{5}{14}\cdot \frac{6}{14}}=0.225$$

\item \textbf{Spearman Distance and Correlation.}
Spearman Distance~\cite{spearman:1904} is the compliment of \emph{Spearman correlation coefficient}, commonly abbreviated as $\rho$. It relates to Pearson distance, however, it is applied to ranked vectors instead of the \emph{PDF}s~\cite{rankCoefficient:1961}.

The ranked vectors can be obtained from the \emph{PDF}s, where numerical values are simply sorted, and non-numerical data is ordered assuming the location implies importance. I.e.,\ for $P= (x, y, z)$, $Q= (z, x, y)$, we get: $rank_P(x)=1$, $rank_P(y)=2$, $rank_P(z)=3$. Thus, $rank_P(P)=(1,2,3)$ and $rank_P(Q) =(3,1,2)$.

Spearman correlation can be used as a measure of a monotonic association for non-normally distributed continuous data. In such cases, Pearson's distance is not accurate enough in describing the core distance between the vectors containing ordinal data or data with outliers. Spearman's correlation is formally defined by:
$$corr_{spr}(P, Q) =1- \frac{ 6\sum_{i=1}^n (rank(P_i)-rank(Q_i))^2}{n(n-1)}$$
Hence, the Spearman distance is formally defined by: 
$$d_{spr}(P, Q) = \frac{ 6\sum_{i=1}^n (rank(P_i)-rank(Q_i))^2}{n(n-1)}$$

For example, using the above $P$,$Q$, we get: $rank(P_1)=1$, $rank(P_2) = 2$, $rank(P_3) =3$, $rank(P_4)=4$ and $rank(Q_1) =1$, $rank(Q_2) =2$, $rank(Q_3) = 3$, $rank(Q_4) = 4$. Thus, $Corr_{spr}(P, Q) = 1- \frac{ 6((1-1)^2 + (2-2)^2 + (3-3)^2 + (4-4)^2)}{12} = 1-0 = 1$ and $d_{spr}(P,Q)=0$.

If the ranked vectors are correlated, the change in the magnitude of one variable in $P$ is associated with a change in the magnitude of  the corresponding variable in $Q$, either in the same  or in the opposite  direction. 
The correlation is stronger if the coefficient approaches an absolute value of $1$. A $0$-coefficient indicates that there is no linear or monotonic association between the vectors. The distance behaves in the opposite direction~\cite{correlation:2018}.

\item \textbf{Squared $\chi^2$ (Triangular Discrimination).}
The Squared $\chi^2$-Distance is a symmetric version of the Pearson Distance. Formally, 
$$d_{sq\chi^2}(P,Q) = \sum^d_{i=1}\frac{(P_i - Q_i)^2}{P_i + Q_i}$$

For example, using the above $P$,$Q$, we get:
\commentout{
$$d_{sq\chi^2}(P,Q) = \frac{(\frac{2}{14}-\frac{1}{14})^2}{\frac{2}{14}+\frac{1}{14}}
+ \frac{(\frac{3}{14}-\frac{2}{14})^2}{\frac{3}{14}+\frac{2}{14}}
+ \frac{(\frac{4}{14}-\frac{5}{14})^2}{\frac{4}{14}+\frac{5}{14}}
+ \frac{(\frac{5}{14}-\frac{6}{14})^2}{\frac{5}{14}+\frac{6}{14}}$$ 
$$=\frac{1}{14^2}\cdot\frac{14}{3}  + \frac{1}{14^2}\cdot\frac{14}{5}  +
              \frac{1}{14^2}\cdot\frac{14}{9}  +\frac{1}{14^2}\cdot\frac{14}{11}  = \frac{364}{14\cdot 495}=0.053$$
}
$$d_{sq\chi^2}(P,Q) = \frac{(\frac{2}{14}-\frac{1}{14})^2}{\frac{2}{14}+\frac{1}{14}}
+ \frac{(\frac{3}{14}-\frac{2}{14})^2}{\frac{3}{14}+\frac{2}{14}}
+ \frac{(\frac{4}{14}-\frac{5}{14})^2}{\frac{4}{14}+\frac{5}{14}}
+ \frac{(\frac{5}{14}-\frac{6}{14})^2}{\frac{5}{14}+\frac{6}{14}}=0.053$$
              
Twice the Squared $\chi^2$-Distance is called the \textbf{Probabilistic Symmetric $\chi^2$}.
%$$d_{Sq\chi} = 2 \sum^d_{i=1}\frac{(P_i - Q_i)^2}{P_i + Q_i}$$

\item \textbf{Divergence Distance.}
The Divergence Distance~\cite{divergence:2001} is obtained by considering the Probabilistic Symmetric $\chi^2$ measure where the denominator is also squared. Formally, 
$$d_{Div}(P,Q) = 2 \sum^d_{i=1}\frac{(P_i - Q_i)^2}{(P_i + Q_i)^2}$$
For example, using the above $P$,$Q$, we get:
\commentout{
$$d_{div}(P,Q) = 2(\frac{(\frac{2}{14}-\frac{1}{14})^2}{(\frac{2}{14}+\frac{1}{14})^2}
+ \frac{(\frac{3}{14}-\frac{2}{14})^2}{(\frac{3}{14}+\frac{2}{14})^2}
+ \frac{(\frac{4}{14}-\frac{5}{14})^2}{(\frac{4}{14}+\frac{5}{14})^2}
+ \frac{(\frac{5}{14}-\frac{6}{14})^2}{(\frac{5}{14}+\frac{6}{14})^2})$$ 
$$= 2(\frac{1}{14^2}\cdot\frac{14^2}{9}  + \frac{1}{14^2}\cdot\frac{14^2}{25}  +
              \frac{1}{14^2}\cdot\frac{14^2}{81}  +\frac{1}{14^2}\cdot\frac{14^2}{121})  = 0.343$$
}
$$d_{div}(P,Q) = 2(\frac{(\frac{2}{14}-\frac{1}{14})^2}{(\frac{2}{14}+\frac{1}{14})^2}
+ \frac{(\frac{3}{14}-\frac{2}{14})^2}{(\frac{3}{14}+\frac{2}{14})^2}
+ \frac{(\frac{4}{14}-\frac{5}{14})^2}{(\frac{4}{14}+\frac{5}{14})^2}
+ \frac{(\frac{5}{14}-\frac{6}{14})^2}{(\frac{5}{14}+\frac{6}{14})^2})= 0.343$$
              
\item \textbf{Clark Distance.}
The Clark Distance~\cite{clark:54} is obtained by taking the squared root of half of the Divergence measure. Formally,
$$d_{Cl}(P,Q) = \sqrt{\sum^d_{i=1}\frac{(P_i - Q_i)^2}{(P_i + Q_i)^2}}$$
For example, using the above $P$,$Q$, we get:
$$d_{Cl}(P,Q) = \sqrt{{364 \over 495}} = 0.857$$
This distance is also known as \textbf{the Coefficient of Divergence}.

\item \textbf{Mahalanobis Distance.}
\cite{Mahalanobis:1930}%, Mahalanobis:1936} 
 An effective multivariate metric that measures the distance between a point and a distribution. It is a multi-dimensional generalization for measuring how many standard deviations away a point $P$ is from the mean of a data-set $D$, being zero for $P$ at the mean vector of $D$ and grows as $P$ moves away from the mean along each principal component axis. In case each of the axes is re-scaled to have unit variance, the Mahalanobis Distance corresponds to the standard Euclidean Distance in the transformed space. It is unitless, scale-invariant, and takes into account the correlations of the data set.

Since Euclidean Distance does not consider how the points in the data-set vary, it cannot be effectively used to estimate how close a point actually is to a distribution of points. If a cluster has an elliptic shape, then some points are closer to the cluster center than others. Yet we cannot conclude that the more distant points, in terms of the classical Euclidean Distance, belong less to the cluster than the closer points.

Let $D$ be a data-set of $n$ $d$-dimensional data points $D_1=(v_{1,1},\ldots,v_{1,d}),\ldots,$ $D_n=(v_{n,1},\ldots,v_{n,d})$ specifying the values of the points in each of the $d$ variables $V_1,\ldots,V_d$. The $d\times d$ matrix $\mathit{cov}(D)$ is the following: 
$$\mathit{cov}(D)=\left(\begin{array}{ccc}
\mathit{cov}(V_1,V_1),& \ldots, & \mathit{cov}(V_1,V_d)\\
\mathit{cov}(V_2,V_1),& \ldots, & \mathit{cov}(V_2,V_d)\\
\vdots & \vdots & \vdots\\
\mathit{cov}(V_d,V_1),& \ldots, & \mathit{cov}(V_d,V_d)\\
\end{array}\right)$$
It represents the shape of the data-set $D$, and measures how far the variables in each dimension are from each other. Naturally, the mean and covariance of the distribution of the data-set are unknown in practice, and hence need to be estimated using the values of $\mu$ and $\mathit{cov}(D)$ in the data sample $D$~\cite{mahanabolis:2018}, thus $\mathit{cov}(V_j,V_{j'})$ is estimated using the formula: 
$$\frac{1}{n-1}\sum_{i=1}^n(v_{i,j}-\mu_j)(v_{i,j'}-\mu_{j'})$$

The Mahalanobis Distance measures the distance from a $d$-dimensional point $P$ to the mean vector $\mu$ of $D$ or between two $d$-dimensional points $P$ and $Q$ using the data-set sample $D$. 

Formally, the Mahalanobis Distance is defined by:
$$d_{\mathit{Mhn}}(P,\mu) =\sqrt{(P-\mu)^T\mathit{cov}(D)^{-1}(P-\mu)}$$
or by
$$d_{\mathit{Mhn}}(P,Q) =\sqrt{(P-Q)^T\mathit{cov}(D)^{-1}(P-Q)}$$

For example~\cite{MahalanobisBlog}, given the sample data-set: 
$$D =\{(64, 580, 29), (66, 570, 33), (68, 590, 37), (69, 660, 46), (73, 600, 55)\},$$ 
where $n=5$, and $d=3$.
We have that $\mu=(68,600,40)$, and the $\mathit{cov}(D)$ matrix is the following:
$$\left(\begin{array}{ccc}
11.5 & 50 & 34.75\\
50 & 1250 & 205\\
34.75 & 205 & 110\\
\end{array}
\right)
$$
The $\mathit{cov}^{-1}(D)$ matrix is the following:
$$\left(\begin{array}{ccc}
3.6885 & 0.0627 & -1.2821\\
0.0627 & 0.0022 & -0.024\\
-1.2821 & -0.024 & 0.4588\\
\end{array}
\right)
$$
Now, given a point $P=(66,640,44)$ we get that: $d_{\mathit{Mhn}}(P,D)=$
$$\sqrt{((66,640,44)-(68,600,40))^T\mathit{cov}(D)^{-1}((66,640,44)-(68,600,40))}=5.33$$
and taking $P=(66,570,33)$ and $Q=(69,660,46)$, we have that: $d_{\mathit{Mhn}}(P,Q)=$
$$\sqrt{((66,570,33)-(69,660,46))^T\mathit{cov}(D)^{-1}((66,570,3)-(69,660,46))}=3.24$$
\end{enumerate}

\subsection{Fidelity Family (Squared-Chord Family)}\label{ss:Fidelity}
\begin{enumerate}
\item \textbf{Fidelity Similarity.}
Bhattacharyya defined a measure of divergence between two multi-nomial populations -- two probability density functions as defined in Subsection~\ref{ss:intersection}.

Let $P=(P_{1},\ldots,P_{d})$ and $Q=(Q_{1},\ldots,Q_{d})$ be two given \emph{PDF}s. Since $\Sigma P_i=1$ and $\Sigma Q_i=1$, then $(\sqrt{P_{1}},\ldots,\sqrt{P_{d}})$ and $(\sqrt{Q_{1}},\ldots,\sqrt{Q_{d}})$ may be considered as the directions of two strait lines through the origin in a $d$-dimensional space. The square of the angle between these two lines is the Bhattacharyya coefficient (denoted as \emph{BC}), also known as Fidelity similarity. All the measures in this family can be expressed using \emph{BC}. Formally~\cite{Bhattacharyya:1943},
$$BC(P,Q)=Sim_{Fid}(P,Q)=\sum^d_{i=1}{\sqrt{P_{i}Q_{i}}}$$
For example, let $P=(\frac{2}{14},\frac{3}{14},\frac{4}{14},\frac{5}{14})$ and $Q=(\frac{1}{14},\frac{2}{14},\frac{5}{14},\frac{6}{14})$, then,
$$BC(P,Q)=Sim_{Fid}(P,Q)=\sqrt{\frac{2\cdot1}{14^2}}+\sqrt{\frac{3\cdot2}{14^2}}+\sqrt{\frac{4\cdot5}{14^2}}+\sqrt{\frac{5\cdot6}{14^2}}=0.986$$

\item \textbf{Bhattacharyya Distance.}
Bhattacharyya~\cite{Bhattacharyya:1943} also defined a distance metric, which is a value in $[0,1]$, providing bounds on Bayes mis-classification probability~\cite{Toussaint:1977}. Formally, 
$$d_{Bha}(P,Q)=-\ln BC(P,Q)= -\ln \sum^d_{i=1}{\sqrt{P_{i}Q_{i}}}$$
For example, using the above $P$,$Q$, we get: $d_{Bha}(P,Q)=-\ln BC(P,Q)=-\ln 0.986=0.014$

\item \textbf{Hellinger Distance.}
It was introduced it in 1909 as a probabilistic analog of the Euclidean Distance. Formally~\cite{Hellinger:1909},
$$d_{Hel}(P,Q)=\frac{1}{\sqrt{2}}\sqrt{\sum^d_{i=1}{(\sqrt P_{i}-\sqrt Q_{i})^{2}}}=\sqrt{1-BC(P,Q)}$$

Also, note that:
$$1-d_{Hel}^2(P,Q)=\sum^d_{i=1}\sqrt{P_{i}Q_{i}}=BC(P,Q)$$

For example, using the above $P$,$Q$, we get: $d_{Hel}(P,Q)=\sqrt{1-0.986}=0.118$

\item \textbf{Matusia Distance.}
Wald introduced a statistical decision theory in 1949~\cite{Wald:1949} in order to deal with decisions made on the basis of observations of a phenomenon obeying probabilistic laws that are not completely known. Matusita formulated a decision's risk~\cite{Matusita:1955}. Formally, $d_{Mat}(P,Q)=$
$$\sqrt{\sum^d_{i=1}{(\sqrt P_{i}-\sqrt Q_{i})^{2}}} = \sqrt{2-2\sum^d_{i=1}{(\sqrt{ P_{i}Q_{i})}}}=\sqrt{2-2\cdot BC(P,Q)}$$

For example, using the above $P$,$Q$, we get: $d_{Mat}(P,Q)=\sqrt{2-2\cdot 0.986}=1.372$

\item \textbf{Squared-Chord Distance.}
The Squared-Chord Distance~\cite{Overpeck:1985,Gavin:2003} is the Matusia Distance omitting the final square root operation. Formally, $$d_{sqc}(P,Q)=\sum^d_{i=1}{(\sqrt P_{i}-\sqrt Q_{i})^{2}}$$

A similarity measure is also defined, by:
$$sim_{sqc}(P,Q)=1-d_{sqc}(P,Q)=2 \sum^d_{i=1}{(\sqrt {P_{i}Q_{i}})}-1= 2\cdot BC(P,Q)-1$$

For example, using the above $P$,$Q$, we get: $d_{sqc}(P,Q)=$
$$(\sqrt{\frac{2}{14}}-\sqrt{\frac{1}{14}})^2+(\sqrt{\frac{3}{14}}-\sqrt{\frac{2}{14}})^2+(\sqrt{\frac{4}{14}}-\sqrt{\frac{5}{14}})^2+(\sqrt{\frac{5}{14}}-\sqrt{\frac{6}{14}})^2=0.026$$
and
$$sim_{sqc}(P,Q)=1-d_{sqc}(P,Q)=0.974$$
\end{enumerate}

\subsection{String Similarity Measures}\label{ss:Strings}
This subsection deals with similarity and distance measures designed for alphabetical based data, which is a common data format in various data science applications. It is split into the following two parts. The first share a common view of a rearrangement process that occurs on the input strings, and the second deals with other views. 

\subsubsection{A. String Rearrangement Measures}
Consider an alphabet-set $\Sigma$ and let $P$ and $Q$ be two $n$-long strings over $\Sigma$. String rearrangement measures involve a process of converting $P$ to $Q$ through a sequence of operations. An {\em operator $\psi$} is a function $\psi: \Sigma^n\rightarrow \Sigma^{n'}$, with the intuitive meaning being that $\psi$ converts $n$-long string $P$ to $n'$-long string $Q$ with a cost associated to $\psi$. For example, an operator $\psi_{k,\ell}$ may be to reverse a sub-string from position $k$ until position $\ell$. Thus $\psi_{2,6}(abrakadabra)=aakarbdabra$. That cost measures the distance between $P$ and $Q$. Formally,

\begin{definition}\label{d:sm} {\rm {\bf [String Rearrangement Metric]}~\cite{AL:10}}
Let $s=(\psi_1,\psi_2,\ldots,\psi_k)$ be a sequence of operators, and let $\psi_s=\psi_1\circ \psi_2 \circ \cdots \circ
\psi_k$ be the composition of the $\psi_j$'s. We say that {\em $s$ converts $P$ into $Q$} if $Q=\psi_s (P)$.

Let $\Psi$ be a set of rearrangement operators, we say that {\em $\Psi$ can convert $P$ to $Q$}, if there exists a sequence $s$ of
operators from $\Psi$ that converts $P$ to $Q$. Given a set $\Psi$ of operators, we associate a non-negative {\em cost}
with each sequence from $\Psi$, $cost:\Psi^*\rightarrow R^+$. The pair $(\Psi,cost)$ is an {\em edit system}.

Given $P,Q\in \Sigma^*$ and an edit system ${\mathcal R}=(\Psi,cost)$, the distance from $P$ to $Q$ under $\mathcal R$ is defined:
\[
d_{\mathcal R}(P,Q)=\min \{ cost(s) | \mbox{$s$ from $\mathcal R$ converts $P$ to $Q$} \}
\]
If there is no sequence that converts $P$ to $Q$ then the distance is $\infty$.
\end{definition}

Note that $d_{\mathcal R}(P,Q)$ is a metric, if there is also the inverse operation with equal cost for each operation in $\Psi$, thus, $d_{\mathcal{R}}(P,Q) = d_{\mathcal{R}}(Q,P)$, and the cost function preserves the triangle inequality. Some of the string rearrangement measures listed below are not metrics, since the triangle inequality may be violated. The cost depends on the rearrangement cost model, which is described next.

\paragraph{\textbf{String Rearrangement Cost Models.}} Several known string rearrangement cost models are used~\cite{AL:10}, where taking each string rearrangement described below with a different cost model gives another string rearrangement variant. The cost models are as follows.
\begin{description}
  \item[\textbf{Unit-Cost Model (UCM):}] In the Unit-Cost Model (UCM) each operation is given a unit cost, so the problem is to transform a given string $P$ into a string $Q$ with a minimum number of operations. For example, using the operator $\psi_{k,\ell}$ defined above, the cost of the operation $\psi_{2,6}(abrakadabra)=aakarbdabra$ is 1.
  \item[\textbf{Length-Cost Model (LCM):}] In the Length-Cost Model (LCM), the cost of an operation depends on its length characteristic, i.e.,\ the difference between the right-most and the left-most positions in the string on which an operation is performed. The cost definition is combined with $L_p$ distance measure (see Subsection~\ref{ss:Minkowski}) defined on the \emph{position differences} instead of the values of the elements in these positions. Such LCM based string metrics are explored in~\cite{AMIR2009359,AmirAILP09}. For example, using the operator $\psi_{k,\ell}$ defined above combined with an $L_2$ distance measure, the cost of the operation $\psi_{2,6}(abrakadabra)=aakarbdabra$ is $(6-1)^2=25$.
  \item[\textbf{Element-Cost Model (ECM):}] In the Element-Cost Model (ECM), the cost of an operation depends on the weights of the elements participating in it. Typically, the cost of an operation is defined to be the sum of the participating elements weights. Exploring ECM is motivated by the observation that some elements may be heavier than other elements. In such cases, moving light elements is preferable to moving heavy elements~\cite{KAPAH20094315}. For example, using the operator $\psi_{k,\ell}$ defined above with the following element-weights: $w(a)=5$, $w(b)=2$, $w(d)=w(k)=w(r)=1$, the cost of the operation $\psi_{2,6}(abrakadabra)=aakarbdabra$ is $2+1+5+1+5=14$, since we moved the elements $b$, $r$, $a$, $k$, $a$ each having its own weight.
\end{description}

Below we describe known string rearrangement measures, where the default model used is UCM.
\begin{enumerate}
\item \textbf{Hamming Distance.}
The Hamming distance measures the similarity of two given equal-length strings by counting the number of positions in which they differ. It is named after the mathematician Richard Hamming, who suggested this measure~\cite{H:1951}. A major application of this measure is in coding theory, more specifically to block codes, in which the equal-length strings are vectors over a finite field. This distance can also be used to naively measure similarity of categorical attributes.

Formally, in Definition~\ref{d:sm}, $\Psi=\{\rho^n_{i,\sigma} | i,n\in \mathbb{N},\ \ i\leq n,\ \ \sigma\in \Sigma\}$, where $\rho^n_{i,\sigma} (\alpha )$ substitutes the $i$th element of $n$-tuple $\alpha$ by symbol $\sigma$.
The Hamming distance is commonly denoted by $H$.

For example, let $P$ be the string $a\ b\ r\ a\  k\ a\ d\ a\ b\ r\ a\ $ and $Q$, the string $b\ a\ r\ a\ k\ a\ d\ a\ b\ r\ a$, then $H(P,Q)=2$, because the first two elements of the strings mismatch so two operators $\rho^{11}_{1,b},\rho^{11}_{2,a}$ should be applied on $P$ in order to get $Q$.

\item \textbf{Edit Distance - Levenshtein Distance.}
The edit distance enables to measure similarity of strings of different length because it enables also to delete and insert characters~\cite{Lev:66}. It is applicable to natural language processing, where automatic spelling correction can determine candidate corrections for a misspelled word by selecting words from a dictionary that have a low distance to the word in question. In bioinformatics, it can be used to quantify the similarity of DNA sequences (see~\cite{Needleman-Wunsch:1970,Smith-Waterman:1981}).

Formally, in Definition~\ref{d:sm}, in addition to the substitution operators of the Hamming distance, $\Psi$ also has {\em insertion} and {\em deletion} operators. The insertion operators are: $\{\iota^n_{i,\sigma} | i,n\in \mathbb{N},\ \ i\leq n,\ \ \sigma\in \Sigma\}$, where $\iota^n_{i,\sigma} (\alpha)$ adds the symbol $\sigma$ following the $i$th element of $n$-tuple $\alpha$, creating an $n+1$-tuple $\alpha'$.

The deletion operators are $\{ \delta^n_{i} | i,n\in \mathbb{N},\ \ i\leq n \}$, where $\delta^n_{i} (\alpha)$ deletes
the symbol at location $i$ of $n$-tuple $\alpha$, creating an $n-1$-tuple $\alpha'$.
The edit distance is commonly denoted by $ED$.

For example, let $P$ be the string $a\ b\ r\ a\ k\ a\ d\ a\ b\ r\ a$ and $Q$, the string $b\ r\ a\ k\ a\ d\ a\ b\ r\ a$, then $ED(P,Q)=1$, because the first element of the string $P$ is deleted using the operator $\delta^{11}_{1}$ in order to get the string $Q$.

\item \textbf{Swap Measure.}
The swap measure is motivated by text editing, where a common typing error is to exchange adjacent string characters. It was suggested by Damerau~\cite{Damerau:1964}.

Formally, in Definition~\ref{d:sm}, $\Psi=\{\zeta^n_{i} | i,n\in \mathbb{N},\ \ i < n \}$, where $\zeta^n_{i} (\alpha )$ swaps the $i$th and $(i+1)$th elements of $n$-tuple $\alpha$, creating an $n$-tuple $\alpha'$. A {\em valid} sequence of operators in the Swap rearrangement has the additional condition that if $\zeta^n_i$ and $\zeta^n_j$ are operators in a sequence then $i\neq j,\ i\neq j+1,\ i\neq j-1$.
Following Definition~\ref{d:sm}, we denote it by $d_{swap}$.

For example, let $P$ be the string $a\ b\ r\ a\ k\ a\ d\ a\ b\ r\ a$ and $Q$, the string $b\ a\ r\ a\ k\ a\ d\ a\ b\ r\ a$, then $d_{swap}(P,Q)=1$, because the first two elements of the string $P$ exchange positions in one swap operation by applying $\zeta^{11}_{1}$ in order to get the string $Q$.
The swap measure is not a metric since the triangle inequality is violated.

\item \textbf{Interchange Distance.}
Unlike the swap operation that allows exchanging only string elements in adjacent positions and to touch a position only once, an interchange operation allows to exchange elements that are apart, while allowing also multiple operations on the same positions. An interchange is a classical distance~\cite{cayley} and generalized by~\cite{ahklpJournal}.

Formally, in Definition~\ref{d:sm}, $\Psi=\{\pi^n_{i,j} | i,n\in \mathbb{N},\ \ i\leq j\leq n \}$, where $\pi^n_{i,j} (\alpha )$ interchanges the $i$th and $j$th elements of $n$-tuple $\alpha$, creating an $n$-tuple $\alpha'$.
Following Definition~\ref{d:sm}, we denote it by $d_{int}$.

For example, let $P$ be the string $a\ b\ r\ a\ k\ a\ d\ a\ b\ r\ a$ and $Q$, the string $a\ k\ r\ b\ a\ a\ d\ a\ b\ r\ a$, then $d_{int}(P,Q)=2$, because the second element of the string $P$ exchanges positions with the fifth element of $P$ using the operator $\pi^{11}_{2,5}$, and then the fifth element of the resulting string $a\ k\ r\ a\ b\ a\ d\ a\ b\ r\ a$ exchanges positions with the forth element in this string using $\pi^{11}_{5,4}$ in order to get the string $Q$.

\item \textbf{Parallel-Interchange Measure.}
The parallel-interchange measure is motivated by rearrangement systems based on the interchange operation, where multiple operations can be performed in parallel~\cite{AMIR2009359}. In order to enable parallelization, a restriction that parallel operations may not touch the same position should be added.

Formally, in Definition~\ref{d:sm}, $\Psi=\{\pi^n_{i,j} | i,n\in \mathbb{N},\ \ i\leq j\leq n \}$, where $\pi^n_{i,j} (\alpha )$  interchanges the $i$th and $j$th elements of $n$-tuple $\alpha$, creating an $n$-tuple $\alpha'$. A {\em valid} sequence of operators in the Parallel-Interchange rearrangement has the additional condition that if $\pi^n_{i,j}$ and $\pi^n_{i',j'}$ are operators in a sequence then $i\neq i',\ j\neq j',\ i\neq j'$ and $i'\neq j$.
Following Definition~\ref{d:sm}, we denote it by $d_{p-int}$.

For example, let $P$ be the string $a\ b\ r\ a\ k\ a\ d\ a\ b\ r\ a$ and $Q$, the string $b\ a\ r\ k\ a\ d\ a\ a\ b\ r\ a$, then $d_{p-int}(P,Q)=2$, because the second element of the string $P$ exchanges positions with the fifth element of $P$ using $\pi^{11}_{2,5}$, and the sixth element of the string $P$ exchanges positions with the seventh element of $P$ using $\pi^{11}_{6,7}$ in order to get the string $Q$. The sequence of these two operations is valid and can be applied in parallel, since the indices are different. The parallel-interchange measure is not a metric since the triangle inequality is violated.
\end{enumerate}

\paragraph{\textbf{Remark:}} The above measures may also be combined, e.g.,\: (1) The well-known \textbf{Damerau-Levenshtein distance} actually combines the Edit and Swap measures~\cite{Damerau:1964}. While the original motivation was to measure distance between human misspellings to improve applications such as spell checkers, Damerau–Levenshtein distance has also seen uses in biology to measure the variation between protein sequences. (2) Another example of measures combinations is suggested by Lipsky et al.~\cite{lppst:2010}, who considered the combined Swap and Hamming measures. 

\subsubsection{B. String Similarity Measures}
Below we describe known string similarity measures that are not based on the string rearrangements model.
\begin{enumerate}
\item \textbf{Longest Common Sub-sequence (LCS).} 
The Longest Common Sub-sequence measure evaluates the similarity of two strings by their shared characters with regard to the characters' order. It is widely used by revision control systems such as Git, data comparison programs such as the \emph{diff} utility, in computational linguistics and biological sequence comparison.

Its strength lies in its simplicity which has allowed development of an extremely fast, bit-parallel computation~\cite{ad:86}. The well-known dynamic programming solution~\cite{Hi-75} requires quadratic running time for LCS computation (for a survey on LCS computation see~\cite{bhr:00}).

Formally, a sub-sequence of a string, is a set of characters that appear in left-to-right order, but not necessarily consecutively. A common sub-sequence of two strings $P,Q$ is a sub-sequence that appears in both strings. A \emph{longest common sub-sequence} of $P,Q$ is a common sub-sequence of maximal length, i.e.,\ $LCS(P,Q)$ is the maximal $\ell$ such that there are $\ell$ characters in $P,Q$, with $P_{i_x}= Q_{j_x}$, for $1\leq x\leq \ell$. $P,Q$ may have different length.

For example, let $P = \textbf{a} \ \textbf{b} \ c \ \textbf{b} \ \textbf{a} \ b \ \textbf{a} \ \textbf{b}$ and let $Q = \textbf{a} \  \textbf{b} \ \textbf{b} \ \textbf{a} \ \textbf{a} \ \textbf{b} \ c \ c$. We have $LCS(P,Q) =6$ and a possible longest common sub-sequence is $a\ b\ b\ a\ a\ b$.

\paragraph{\textbf{Relation between LCS and Edit Distance:}} For two strings $P$ of length $m$ and $Q$ of length $n$, the edit distance when only insertion and deletion are allowed (no substitution), or when the cost of the substitution is the double of the cost of an insertion or deletion, is:
$$ED(P,Q)=n+m-2\cdot \left|LCS(P,Q)\right|.$$

\paragraph{\textbf{LCS Refinements:}} The disadvantage of the LCS is that it might be a crude similarity measure because consecutive matching letters in the LCS can have different spacing in the two sequences, i.e.,\ there is no penalty for insertion and deletion and no limitations on their ``distribution''. Consider for example the sequences: $P = (a b a)^{n/3}$ and $Q = (b c c )^{n/3}$, for which the LCS, $b^{n/3}$ is quite large, but there are no two matched symbols consecutive in both sequences. Any common sub-sequence of this size ``put together'' separated elements implying a rather ``artificial'' similarity. Below we detail some variants of LCS aiming at refining the measure for certain scenarios:
\begin{enumerate}
\item \textbf{Longest Common $k$-Length Sequences (LCSk).}
Benson et al.~\cite{bls:2016} suggested the Longest Common $k$-Length Sequences (LCSk), where the common sub-sequence is required to consist of at least k length sub-strings. Formally, let $P,Q$ be two $n$-length strings over alphabet $\Sigma $, $LCSk(P,Q)$ is the maximal $\ell$ such that there are $\ell$ sub-strings, $P_{i_x}, \ldots,$ $P_{i_x+k-1} = Q_{j_x},\ldots, Q_{j_x+k-1}$ for $1\leq x\leq \ell $, where $i_x + k - 1< i_{x+1}$ and $j_x + k - 1 < j_{x+1}$. $LCS(P,Q)=LCS1(P,Q)$.

For example, let $P = a \ b \ c \ \textbf{b} \ \textbf{a} \  b \  \textbf{a} \ \textbf{b}$ and let $Q = \textbf{b} \ \textbf{a} \  a \  b \  \textbf{a} \ \textbf{b} \ c \ c$. We have $LCS2(P,Q) = 2$ and a possible longest common 2-length sub-sequence is $b\ a\ a\ b$, while $LCS3(P,Q) = 1$, and a possible longest common 3-length sub-sequence is $a\ b\ c$ achieved by matching positions 1,2,3 in $P$ with positions 5,6,7 in $Q$.

\item \textbf{Heaviest Common Sub-sequence (HCS).}
Jacobson and Vo~\cite{JV:1992} suggested the Heaviest Common Sub-sequence measure, where each alphabet character in $\Sigma$ has a positive weight assigned to it. The weight of a string over $\Sigma$ is defined as the sum of the weights of the characters in it.\commentout{ The HCS measures takes also the weights into account and tries to maximize the total weight of the common sub-sequence.} Formally, let $P$ and $Q$ be two strings over an alphabet $\Sigma$ with weight function $w:\Sigma \rightarrow \mathbb{R}$. The heaviest common sub-sequence of $P$ and $Q$ is the length of a sequence $Z$ where $Z$ is a common sub-sequence for $P$ and $Q$ maximizing $\sum_{z_i\in Z} w(z_i)$. The heaviest common sub-sequence of $P$ and $Q$ is commonly denoted by $HCS(P,Q)$.

For example, let $P = a \ b \ c \ b \ a \ b \ a \ b$, $Q = c \ a \ b \ b \ a \ a \ b \ c $, and assume that the weights of the characters are: $w(a)=1$, $w(b)=1$, $w(c)=4$. We have $HCS(P,Q)=8$ and a possible common sub-sequence having this weight is $c\ b\ a\ a\ b$ of length 5. Note that $LCS(P,Q)=6$, where a longest common sub-sequence is $a \ b \ b \ a \ a \ b$ with weight 6.
\end{enumerate}

\paragraph{\textbf{LCS Generalizations and Variants:}} The LCS problem has been investigated on more general structures such as trees and matrices~\cite{ahkst:08}, run-length encoded strings~\cite{als:99}, alternative definitions for weighted sub-sequences~\cite{ags:10} and variants of the problem, such as LCS alignment~\cite{lz:01}\commentout{, constrained LCS~\cite{cc:11}}, restricted LCS~\cite{ghll:10} and LCS approximation~\cite{lln:11}.

\item \textbf{Jaro Similarity.}
Inspired by edit distance and motivated by the \emph{record linkage problem}, i.e.,\ the task of finding records in a data set that refer to the same entity across different data sources, it counts the number of matches (unlike the Hamming distance that counts mismatches), however, widening the view of a match between characters to include also those that do not appear in the same position in both strings while heuristically limiting how far such characters can be. It gives a value in $[0,1]$, where 0 means extreme dissimilarity and 1 means extreme similarity (not necessarily identity).

Formally, let $P$ and $Q$ be two strings over an alphabet $\Sigma$, and let $|P|$, $|Q|$ be the length of $P$ and $Q$, respectively. Define $m$ to be the number of matching characters in $P$ and $Q$, where two characters are considered matching only if they are the same in $P$ and $Q$ and not farther than $\lfloor\frac{\max\{|P|,|Q|\}}{2}\rfloor-1$ characters apart. Define $t$ (transpositions) to be the number of matching characters (but different position) divided by 2. The Jaro similarity is~\cite{Jaro:1989}:
$$sim_{\mathit{Jaro}}(P,Q)= \left\{
  \begin{array}{ll}
    0, & \hbox{if m=0;} \\
    \frac{1}{3}(\frac{m}{|P|}+\frac{m}{|Q|}+\frac{m-t}{m}), & \hbox{otherwise.}
  \end{array}
\right.$$

For example, let $P = a \ b \ c \ b \ a \ b \ a \ b$, $Q = c \ a \ b \ b \ a \ a \ b \ c $, we have that $|P|=|Q|=8$, $\lfloor\frac{\max\{|P|,|Q|\}}{2}\rfloor-1=3$, thus, matching characters are allowed to be at most 3 characters apart, therefore, $m=7$, $t=5/2=2.5$, since out of 7 matching characters only 2 were in the same positions in $P$ and $Q$, so: 
$$sim_{\mathit{Jaro}}(P,Q)=\frac{1}{3}(\frac{7}{8}+\frac{7}{8}+\frac{7-2.5}{7})=\frac{2.3929}{3}=0.7976$$

The Jaro distance $d_{\mathit{Jaro}}$ is defined as: $$d_{\mathit{Jaro}}(P,Q)=1-sim_{\mathit{Jaro}}(P,Q)$$

The Jaro distance is not a metric because it neither obeys the triangle inequality nor satisfies the identity axiom requiring that $d(P,Q)=0$ if and only if $P=Q$.

The following is a known variant of the Jaro similarity:
\begin{enumerate}
\item \textbf{Jaro–Winkler Similarity.}
Winkler’s rationale was that people entering data are less likely to make mistakes in the first 4 characters and these are more likely to be noticed and corrected. As such, the Jaro-Winkler similarity is more favoring on matching prefixes in the first 4 characters.

It uses two parameters: a predefined scaling factor $p$ and $\ell$, which is the minimum between 4 and the length of the longest common prefix of the two strings. The scaling factor $p$ specifies how much the score is rewarded for having common prefixes (or, in the distance version, penalized for having mismatches in the length 4 prefix), and should not exceed 0.25 with 4 being the maximum prefix length considered, otherwise the similarity could become larger than 1. It standard value is $p=0.1$.

Formally, let $P$ and $Q$ be two strings over an alphabet $\Sigma$, and let $|P|$, $|Q|$ be the length of $P$ and $Q$, respectively.
The Jaro–Winkler similarity is defined~\cite{winkler:1990} as follows:
$$sim_{\mathit{JW}}(P,Q)=sim_{\mathit{Jaro}}(P,Q)+\ell\cdot p(1-sim_{\mathit{Jaro}}(P,Q))$$

For example, let $P = a \ b \ c \ b \ a \ b \ a \ b$, $Q = c \ a \ b \ b \ a \ a \ b \ c $, as above, we have that $sim_{\mathit{Jaro}}(P,Q)=0.7976$, thus, if 4 is the maximum considered prefix length, and $p=0.1$, we have that $\ell=0$, since $P$ and $Q$ have no common prefix, so, in this case:
$$sim_{\mathit{JW}}(P,Q)=sim_{\mathit{Jaro}}(P,Q)=0.7976$$
However, considering $P = a \ b \ c \ b \ a \ b \ a \ b$, $Q = a \ b \ b \ a \ a \ b \ c \ c$, as before, we have that $|P|=|Q|=8$, thus, matching characters are allowed to be at most 3 characters apart, but in this case, $m=6$, $t=1/2=0.5$, so $$sim_{\mathit{Jaro}}(P,Q)=\frac{1}{3}(\frac{6}{8}+\frac{6}{8}+\frac{6-0.5}{6})=\frac{2.4167}{3}=0.8056$$
while, $\ell=2$, so: $sim_{\mathit{JW}}(P,Q)=0.8056+2\cdot 0.1(1-0.8056)=0.8445$.
\commentout{
$$sim_{\mathit{JW}}(P,Q)=0.8056+2\cdot 0.1(1-0.8056)=0.8056+0.03888=0.8445$$
}
The Jaro–Winkler distance $d_{JW}$ is defined as: 
$$d_{JW}(P,Q)=1-sim_{JW}(P,Q)$$

Note, that if $p=0.25$ is selected, any string pair with the same first 4 letters will have a Jaro-Winkler similarity of 1 and a Jaro-Winkler distance of 0. The Jaro–Winkler distance is not a metric because it neither does not obey the triangle inequality nor satisfy the identity axiom requiring that $d(P,Q)=0$ if and only if $P=Q$.
\end{enumerate}

\item \textbf{N-Grams Measure.}\footnote{Sometimes called Q-grams.}
\commentout{This similarity/distance measure allows to determine the length of the sub-strings ($N$-grams) compared. }N-grams compare sub-strings of $N$ consecutive characters, where $N$ can range from 1 to the string length. For example, bi-grams are $N$-grams of length 2 (e.g., “ab”), tri-grams are $N$-grams of length 3 (e.g., “abc”), etc. The $N$-gram distance measure compares all possible $N$-grams of the strings and returns the number of unpaired $N$-grams between two sets of strings.\commentout{ Thus, a value of 0 means there are no unpaired strings and any value above that is the number of unpaired strings.} Formally, let $P$ and $Q$ be two strings over an alphabet $\Sigma$, denote the set of $N$-grams of $P$ by $\mathit{Ngrams}(P)$ and the set of $N$-grams of $Q$ by $\mathit{Ngrams}(Q)$, then the $N$-gram similarity and distance measures are obtained~\cite{kondrak2005n,Alberto:2010} by the following formulas:
$$sim_{\mathit{Ngrams}}(P,Q)=|\mathit{Ngrams}(P)\cap \mathit{Ngrams}(Q)|$$
$$d_{\mathit{Ngrams}}(P,Q)=|\mathit{Ngrams}(P)\cup \mathit{Ngrams}(Q)|-|\mathit{Ngrams}(P)\cap \mathit{Ngrams}(Q)|$$

For example, let $N=2$ and let $P = a \ b \ c \ b \ a \ b \ a \ b$, $Q = a \ b \ b \ a \ a \ b \ c \ c$, then the set of 2-grams of $P$ is: $\mathit{2grams}(P)=\{\textbf{ab}, \textbf{bc}, cb, \textbf{ba}\}$ and the set of 2grams of $Q$ is $\mathit{2grams}(Q)=\{\textbf{ab}, bb, \textbf{ba}, aa, \textbf{bc}, cc\}$. Therefore,
$$sim_{\mathit{2grams}}(P,Q)=3 \    \ \rm{and}  \  \ d_{\mathit{2grams}}(P,Q)=7-3=4$$

The $N$-gram distance is not a metric because it neither obey the triangle inequality nor satisfy the identity axiom requiring that $d(P,Q)=0$ if and only if $P=Q$.

\paragraph{\textbf{N-Grams Based Measures:}} There are several common uses of other similarity/distance measures that are defined on sets or vectors applied to strings based on the pre-computation of the $N$-grams. Below we describe two of them:
\begin{enumerate}
\item \textbf{$N$-Grams Jaccard Index.}
The Jaccard distance on the $N$-grams sets is similar to the $N$-grams distance, but calculates the number of shared $N$-grams between two strings (the intersection) divided by the union of all $N$-grams in the two strings. A perfect match returns a score of zero, while no shared $N$-grams returns a score of 1.

Formally, let $P$ and $Q$ be two strings over an alphabet $\Sigma$, denote the set of $N$-grams of $P$ by $\mathit{Ngrams}(P)$ and the set of $N$-grams of $Q$ by $\mathit{Ngrams}(Q)$, then the Jaccard similarity and distance defined on the $N$-grams sets are:
$$sim_{\mathit{JN-N}}(P,Q)=\frac{|\mathit{Ngrams}(P)\cap \mathit{Ngrams}(Q)|}{|\mathit{Ngrams}(P)\cup \mathit{Ngrams}(Q)|}$$
%$$d_{\mathit{JN-N}}(P,Q) = 1-sim_{\mathit{JN-N}}(P,Q)=\frac{|\mathit{Ngrams}(P)\cup \mathit{Ngrams}(Q)|-|\mathit{Ngrams}(P)\cap \mathit{Ngrams}(Q)|}{|\mathit{Ngrams}(P)\cup \mathit{Ngrams}(Q)|}$$
\begin{eqnarray}
% \nonumber to remove numbering (before each equation)
  \nonumber d_{\mathit{JN-N}}(P,Q) &=& 1-sim_{\mathit{JN-N}}(P,Q)= \\
  \nonumber &=& \frac{|\mathit{Ngrams}(P)\cup \mathit{Ngrams}(Q)|-|\mathit{Ngrams}(P)\cap \mathit{Ngrams}(Q)|}{|\mathit{Ngrams}(P)\cup \mathit{Ngrams}(Q)|}
\end{eqnarray}

For example, let $N=2$ and let $P = a \ b \ c \ b \ a \ b \ a \ b$, $Q = a \ b \ b \ a \ a \ b \ c \ c$, then the set of 2-grams of $P$ is: $\mathit{2grams}(P)=\{\textbf{ab}, \textbf{bc}, cb, \textbf{ba}\}$ and the set of 2grams of $Q$ is $\mathit{2grams}(Q)=\{\textbf{ab}, bb, \textbf{ba}, aa, \textbf{bc}, cc\}$. Therefore,
$$sim_{\mathit{JN-2}}(P,Q)=\frac{3}{7}=0.4286 \    \ \rm{and}  \  \ d_{\mathit{JN-2}}(P,Q)=\frac{7-3}{7}=0.5714$$

\item \textbf{N-Grams Cosine Distance.}
The Cosine measure captures the angle between two vectors rather than differences in attributes. When applied to $N$-grams, the measure takes all of the $N$-grams from each string, sorts them alphabetically into unique values, then counts occurrences of each $N-gram$ in the original strings to create vectors. The angles of the vectors are calculated and then subtracted from 1.

Formally, let $P$ and $Q$ be two strings over an alphabet $\Sigma$, denote the set of q-grams of $P$ by $\mathit{Ngrams}(P)$ and the set of $N$-grams of $Q$ by $\mathit{Ngrams}(Q)$. Let $S=\mathit{Ngrams}(P)\cup \mathit{Ngrams}(Q)$ sorted by alphabetic order, where $s$ is the size of $S$, and let $P^{(S)},Q^{(S)}$ be the vectors generated for $P$ and $Q$ respectively, by setting, for each $i$, $1\leq i\leq s$, $P^{(S)}[i]$, $Q^{(S)}[i]$ to be the number of times the $i$-th $N$-gram in $S$ appear in $P$, $Q$, respectively, then, the Cosine similarity and distance on the $N$-grams sets are as follows:
%$$sim_{\mathit{CosN-N}}(P,Q)=sim_{\mathit{Cos}}(P^{(S)},Q^{(S)})=\frac{\sum_{i=1}^{s}P^{(S)}[i]\cdot Q^{(S)}[i]}{\sqrt{\sum_{i=1}^{s}P^{(S)}[i]^2}\cdot \sqrt{\sum_{i=1}^{s}Q^{(S)}[i]^2}}$$
\begin{eqnarray}
% \nonumber to remove numbering (before each equation)
  \nonumber sim_{\mathit{CosN-N}}(P,Q) &=& sim_{\mathit{Cos}}(P^{(S)},Q^{(S)})= \\
  \nonumber &=& \frac{\sum_{i=1}^{s}P^{(S)}[i]\cdot Q^{(S)}[i]}{\sqrt{\sum_{i=1}^{s}P^{(S)}[i]^2}\cdot \sqrt{\sum_{i=1}^{s}Q^{(S)}[i]^2}}
\end{eqnarray}
$$d_{\mathit{CosN-N}}(P,Q)=d_{\mathit{Cos}}(P^{(S)},Q^{(S)})=1-sim_{\mathit{Cos}}(P^{(S)},Q^{(S)})$$

For example, let $N=2$ and let $P = a \ b \ c \ b \ a \ b \ a \ b$, $Q = a \ b \ b \ a \ a \ b \ c \ c$, then the set of 2-grams of $P$ is: $\mathit{2grams}(P)=\{ab, bc, cb, ba\}$ and the set of 2grams of $Q$ is $\mathit{2grams}(Q)=\{ab, bb, ba, aa, bc, cc\}$. Thus, $S=\{aa, ab, ba, bb, bc,cb,cc\}$, $P^{(S)}=(0,3,2,0,1,1,0)$, $Q^{(S)}=(1,2,1,1,1,0,1)$. Therefore,
\commentout{
$$sim_{\mathit{CosN-N}}(P,Q)=\frac{9}{\sqrt{15}\cdot\sqrt{9}}=\frac{9}{11.619}=0.7746$$
}
$$sim_{\mathit{CosN-N}}(P,Q)=\frac{9}{\sqrt{15}\cdot\sqrt{9}}=0.7746 \    \ \rm{and}  \  \ d_{\mathit{CosN-N}}(P,Q)=0.2254$$
\end{enumerate}
\end{enumerate}

\section{Conclusion}\label{s:conclusion}
In this guide we described a comprehensive set of prevalent similarity measures. The necessity of such a guide to the research community is its easing the task of choosing a proper similarity measure for a given scenario and understanding the principles of designing similarity measures.     

%\begin{acknowledgements}
%If you'd like to thank anyone, place your comments here
%and remove the percent signs.
%\end{acknowledgements}

% BibTeX users please use one of
% basic style, author-year citations
%\bibliographystyle{ACM-Reference-Format}
%\bibliographystyle{plain}

%\bibliography{similarity}

\end{document}